\documentclass{emulateapj} 
\usepackage{graphicx}
\usepackage{amsmath}   
\usepackage{float}
\usepackage{url}
\usepackage{color}
\usepackage{amssymb }
\usepackage{epsfig}
\usepackage{epstopdf}

\graphicspath{{./}}
\usepackage{wasysym}

\usepackage{natbib}

\allowdisplaybreaks[1]

\newcommand{\Msun}{\rm{M_\odot}}
\newcommand{\Rsun}{\rm{R_\odot}}

\newcommand{\Rearth}{\rm{R_\oplus}}
\newcommand{\pc}{\rm {pc}}
\newcommand{\au}{\rm{AU}}

\newcommand{\teff}{T_{\rm eff}}
\newcommand{\teq}{\rm{T_{\rm eq}}}
\newcommand{\binary}{{\it Kepler\rm -296}}
\newcommand{\triple}{{\rm KOI-2626}}
\newcommand{\newbin}{{\rm KOI-3049}}
\newcommand{\chisq}{$\chi^2$}
\newcommand{\ebv}{{\rm E}(\bv)}
\newcommand{\kepler}{\it Kepler \rm}
\newcommand{\pid}{GO-12893}

\newcommand{\binma}{0.626}
\newcommand{\binmaerr}{0.082}
\newcommand{\binmb}{0.453}
\newcommand{\binmberr}{0.082}
\newcommand{\binta}{3821}
\newcommand{\bintaerr}{160}
\newcommand{\bintb}{3434}
\newcommand{\bintberr}{156}
\newcommand{\binra}{0.595}
\newcommand{\binraerr}{0.072}
\newcommand{\binrb}{0.429}
\newcommand{\binrberr}{0.072}
\newcommand{\binchisq}{0.218}
\newcommand{\bind}{360}
\newcommand{\binderr}{20}
\newcommand{\bina}{80}
\newcommand{\binaerr}{5}
\newcommand{\binp}{660}
\newcommand{\binperr}{60}
\newcommand{\tripma}{0.501}
\newcommand{\tripmaerr}{0.086}
\newcommand{\tripmb}{0.436}
\newcommand{\tripmberr}{0.086}
\newcommand{\tripmc}{0.329}
\newcommand{\tripmcerr}{0.085}
\newcommand{\tripta}{3649}
\newcommand{\triptaerr}{166}
\newcommand{\triptb}{3523}
\newcommand{\triptberr}{160}
\newcommand{\triptc}{3391}
\newcommand{\triptcerr}{158}
\newcommand{\tripra}{0.478}
\newcommand{\tripraerr}{0.075}
\newcommand{\triprb}{0.415}
\newcommand{\triprberr}{0.077}
\newcommand{\triprc}{0.321}
\newcommand{\triprcerr}{0.076}
\newcommand{\tripchisq}{0.860}
\newcommand{\tripd}{340}
\newcommand{\tripderr}{35}
\newcommand{\tripab}{70}
\newcommand{\tripaberr}{7}
\newcommand{\tripac}{55}
\newcommand{\tripacerr}{6}
\newcommand{\newbinma}{0.607}
\newcommand{\newbinmaerr}{0.081}
\newcommand{\newbinmb}{0.557}
\newcommand{\newbinmberr}{0.081}
\newcommand{\newbinta}{4529}
\newcommand{\newbintaerr}{163}
\newcommand{\newbintb}{4274}
\newcommand{\newbintberr}{159}
\newcommand{\newbinra}{0.588}
\newcommand{\newbinraerr}{0.071}
\newcommand{\newbinrb}{0.536}
\newcommand{\newbinrberr}{0.071}
\newcommand{\newbinchisq}{0.907}
\newcommand{\newbind}{485}
\newcommand{\newbinderr}{20}
\newcommand{\newbina}{225}
\newcommand{\newbinaerr}{10}
\newcommand{\newbinp}{3150}
\newcommand{\newbinperr}{205}

\begin{document}

\title{Revision of Earth-sized {\it Kepler} Planet Candidate Properties with High Resolution Imaging by {\it Hubble Space Telescope}\footnotemark[*]}
\author{Kimberly~M.~S.~Cartier\altaffilmark{1,2}, 
{Ronald~L.~Gilliland}\altaffilmark{1,2}, 
{Jason~T.~Wright}\altaffilmark{1,2}, and 
{David~R.~Ciardi}\altaffilmark{3}}	
\altaffiltext{1}{Department of Astronomy \& Astrophysics, The Pennsylvania State University, 525 Davey Lab, University Park, PA 16802}
\altaffiltext{2}{Center for Exoplanets and Habitable Worlds, The Pennsylvania State University, University Park, PA 16802}
\altaffiltext{3}{NASA Exoplanet Science Institute, California Institute of Technology, Pasadena, CA, USA}
\email{kms648@psu.edu} 
\footnotetext[*]{Based on observations with the NASA/ESA {\it Hubble Space Telescope}, obtained at the Space Telescope Science Institute, which is operated by AURA, Inc., under NASA contract NAS 5-26555.}

\shortauthors{Cartier~et~al.}
\shorttitle{Revision of \kepler Planet Candidates with {\it HST}}

\slugcomment{Accepted version March, 5, 2015, by ApJ }

\begin{abstract}
 
We present the results of our {\it Hubble Space Telescope} program and describe how our analysis methods were used to re-evaluate the habitability of some of the most interesting \kepler planet candidates. Our program observed 22 {\kepler Object of Interest (KOI) host stars}, several of which were found to be multiple star systems unresolved by {\it Kepler}. We use our high-resolution imaging to spatially resolve the stellar multiplicity of \binary, \triple, and \newbin, and {develop} a conversion to the \kepler photometry (Kp) from the F555W and F775W filters on WFC3/UVIS. The binary system \binary\,{(5 planets)} has a projected separation of $0\farcs217$ (\bina\,\au); \triple\,{(1 planet candidate)} is a triple star system with a projected separation of $0\farcs201$ (\tripab\,\au) between the primary and secondary components and $0\farcs161$ (\tripac\,\au) between the primary and tertiary; and the binary system \newbin\,{(1 planet candidate)} has a projected separation of $0\farcs464$ (\newbina\,\au). {We use our measured photometry to fit the separated stellar components to the latest Victoria-Regina Stellar Models with synthetic photometry to {conclude} that the systems are coeval.} The components of the three systems range from mid-K dwarf to mid-M dwarf spectral types.{We solved for the planetary properties of each system analytically and via an MCMC algorithm using our independent stellar parameters}. The planets range from {$\rm{\sim{1.6}\,\Rearth\,to\,\sim4.2\,\Rearth}$, mostly Super Earths and mini-Neptunes}. As a result of the stellar multiplicity, some planets previously in the Habitable Zone are, in fact, not, and other planets may be habitable depending on their assumed stellar host.
\end{abstract}
\section{Introduction}
\label{sec:intro}

\setcounter{footnote}{0}
{Since its advent, the \kepler mission has increased the number of candidate exoplanets by thousands, confirmed hundreds of planets, and has pushed the boundaries of transiting exoplanets to smaller radii and longer orbital periods than previously detected} \citep{batalha13, borucki10, borucki11, burke14, fressin13, howard12, liss14}. The 2013 release of the first 16 quarters of \kepler data has increased the number of known {transiting} exoplanet candidates of all radii, but has been especially fruitful for the smallest candidates (with a fractional increase of 201\% known planets smaller than $2\,R_\oplus$) and for the longest orbital periods \citep[with a fractional increase of $124\%$ for orbits longer than 50 days;][]{batalha13, borucki11}. More recently, \cite{rowe14} has nearly doubled the total number of validated exoplanets through careful elimination of false-positive detections in multi-planet systems. {Nearly{40\%} of \kepler planet candidates have been found to reside in multiple planet systems \citep{batalha13, rowe14} }and recent surveys show that the vast majority of multiple transiting {system detections} are true multiple planet systems \citep{liss14,rowe14}. {\cite{howard12}} showed that the planet occurrence rate increases from F to K dwarfs, and followup studies by \cite{dres13} and \cite{dresrevise} {showed} that this trend continues increasing towards M dwarfs. New estimates of $\eta_\oplus$ have made use of these more robust {data}, arriving at a conservative prediction that between {6-15\%} of Sun-like stars have an Earth-size planet in the Habitable Zone \citep[HZ;][]{kasthz, petigura13, silburt14}, though utilization of state-of-the-art {Habitable Zone  calculations} will likely reduce this number \citep{ravihz}.

While the majority ($>2000$) of the \kepler planet candidates reside in apparently single-star systems, this percentage is likely due to a selection effect that avoids binary targets \citep{kratter12}. Accounting for the frequency of binary stars, the occurrence of planets in multiple star systems could be as high as 50\% \citep{kaib13}. Nearly all of the \kepler targets have been imaged by the United Kingdom Infrared Telescope (UKIRT) or other ground based telescopes that provide $\sim1\arcsec$ seeing,{but only 30.5\%\footnote{\url{http://cfop.ipac.caltech.edu/home/}} }of planet candidate hosts have been followed up with speckle interferometry, adaptive optics imaging, or other high-resolution imaging capable of resolving tightly bound systems. This implies that a significant fraction of \kepler targets may in fact be close-in binary or higher multiple star systems that remain unresolved. Recent advancements in ground-based adaptive optics (AO), particularly at the {Keck} Observatory, have {accelerated} high-resolution imaging of \kepler Objects of Interest (KOIs), {especially} those with the smallest planets at the coolest temperatures. The identification of any diluting sources in the aperture allows for improved precision when determining planet habitability and can also reveal previously unresolved stellar {companions}. \cite{gill11} and \cite{gs14} have shown that the {sharp and stable point spread function (PSF) of the WFC3 camera on {\it Hubble Space Telescope} is} ideal for detailed photometric study of \kepler targets and for the identification of field stars in the {\it HST} photometric aperture down to about $\Delta \rm{mag}=10$. The F555W and F775W filters on WFC3/UVIS are ideally suited to observe the majority of \kepler targets.

Our {\it HST} Guest Observing {Snapshot} Program \pid\, observed 22 targets before May 1, 2014, six of which were found to be multiple star systems unresolved by {\it Kepler}. \cite{gs14} discusses the overarching scientific goals and conclusions of {the observing program},  including program parameters and basic image analysis, stellar companion detections and detection completeness, comparison to other high-resolution imaging, and tests for physical association of detected stellar companions.  \cite{gs14} presents analysis that directly supports the methods in this paper, and serves as a companion paper to this work. {Here, we perform multiple-star isochrone fitting} using the latest release of the Victoria-Regina Stellar Models \citep{vdb, vdb2} for three \kepler targets of particular interest: KIC 11497958 (KOI-1422, hereafter \binary), KIC 11768142 (hereafter \triple), and KIC 6263593 (hereafter \newbin). We discuss the parameters of \pid\, and our image analysis in Section \ref{sec:obs}, including our use of the DrizzlePac software and our conversion of our {\it HST} photometry to the \kepler photometric bandpass. In Section \ref{sec:multistar} we discuss the importance of our three targets and detail our characterization of the stellar components in each multi-star system, including the use of {our empirically derived PSF} to calculate the photometry of our systems, fitting to the Victoria-Regina isochrones, and examination of their suitability for our targets. Section \ref{sec:planets} presents our re-evaluation of the planetary habitability. For the purposes of this paper, we define a ``habitable planet" to be a planet that falls between the moist greenhouse limit and the maximum greenhouse limit as defined by \cite{ravihz}. Finally, we discuss our results in context of previous and future work in Section \ref{sec:discussion} and summarize our findings in Section \ref{conclusion}.
\section{Observations and Image Analysis}
\label{sec:obs}
The 158 targets proposed for {observation} were selected from the 2013 data release of \kepler planet candidates by \cite{batalha13}, prioritized by smaller candidate radius and cooler equilibrium temperature. The remaining ranked targets were then sorted between ground-based AO and {\it HST} observations based on the quality of observations for the fainter targets, where {\it HST} would provide comparable or better data in half an orbit than a full night of ground-based AO observation on Lick or Palomar systems. This resulted in the selected {\it HST} targets having the shallowest transit signatures, which thus require the deepest imaging. The targets have a nominal upper limit of $R_p< 2.5\Rearth$ \citep{batalha13},{though our revision of the stellar parameters indicates that some of the planets are actually larger than this limit}. Of the 158 proposed targets, 22 were observed before May 2014 and are included in our analysis. Any observations collected after May 2014 will be analyzed using the techniques presented {in this section}, but are not included in this paper. Our image analysis utilized the latest image registration and drizzling software from STScI DrizzlePac \citep{drizzlepac} and our own PSF definition and subtraction.


\subsection{{\it HST} High Resolution Imaging}
\label{sec:highres}
Our \it HST \rm  program provided high resolution imaging in the F555W ($\lambda\sim0.531\micron$) and F775W ($\lambda\sim0.765\micron$) filters of the WFC3/UVIS camera to support the analysis of faint KOIs. In particular, the parameters of our observations allowed us to examine the properties of faint stellar hosts of small and cool planet candidates. At the faint magnitudes of typical \kepler stars, our WFC3 imaging provides resolution that is competitive with current ground-based AO and has the advantage of using two well calibrated optical filters well matched to the \kepler bandpass. 

The observations made by {\emph{HST}} closely resemble those made by \cite{gill11}, though we only used observations in F555W and F775W since the faintest \kepler targets could still be probed in these bandpasses. Observations planned for each of the 158 SNAP targets were identical in form. In each filter, we took 5 observations of each target: 4 observations with exposure times to reach 90\% of full well depth in the brightest pixel, and an additional observation at an exposure time equal to 50\% more than the sum of the unsaturated exposures to bring up the wings of the PSF. The saturated exposure yielded a $\Delta$-mag of $\sim9$ outside $2\arcsec$ and helped with the signal-to-noise anywhere outside the inner $0\farcs1$.

\subsection{AstroDrizzle}
\label{sec:astrodrizzle}


\begin{figure}[t]
	\begin{center}
	\includegraphics[width=0.4\textwidth]{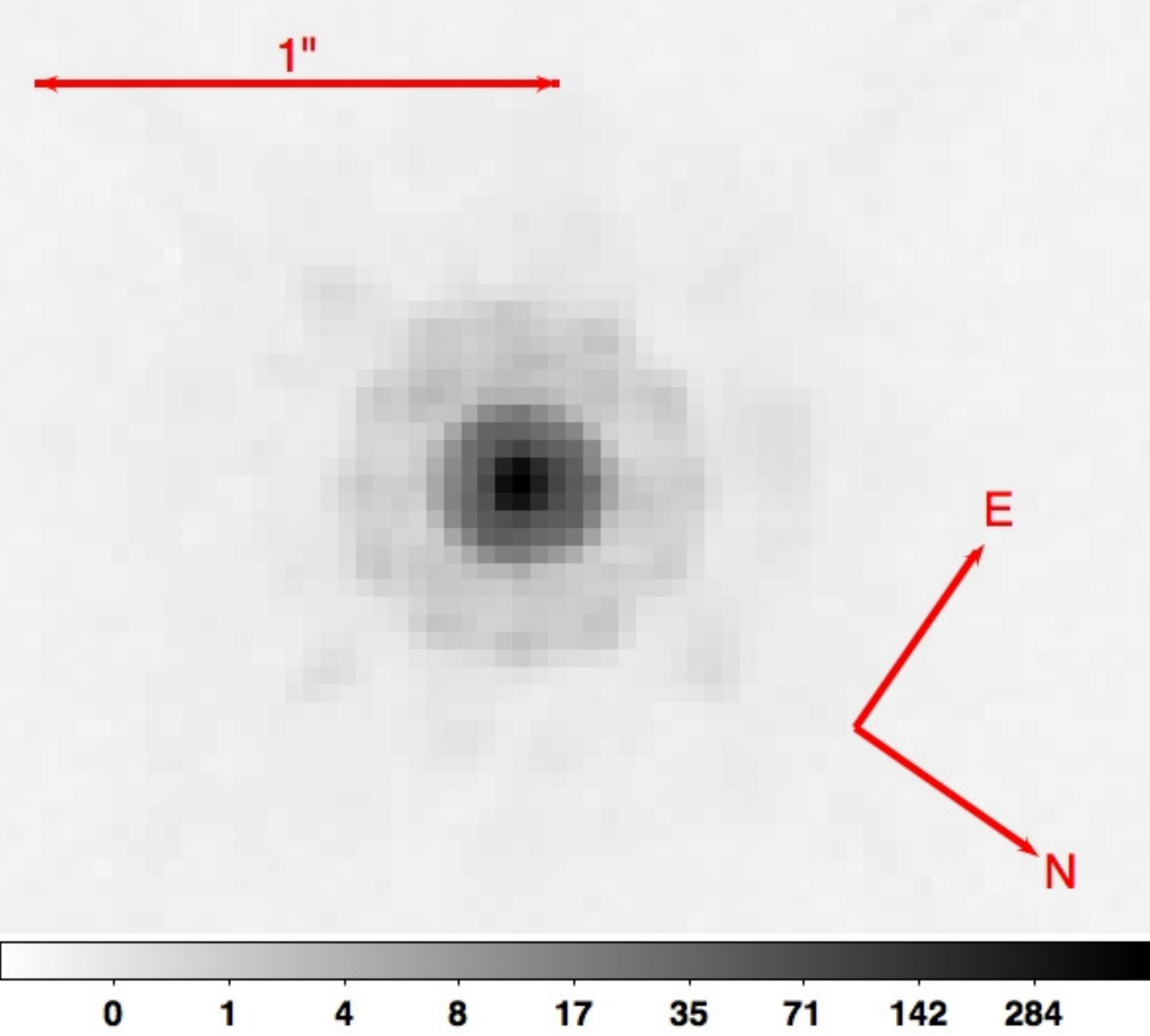}
	\caption{AstroDrizzled image of KIC 4139816 in the F775W filter showing a $1\farcs0$ scale bar and orientation. The image is approximately $2\farcs0$ on a side. Units are $\log_{10}$ of $e^-/s$. The FWHM of the PSF is $0\farcs0777$.}     
	\label{fig:775driz}
	\end{center}
\end{figure}

{The ``drizzle" process, formally known as variable-pixel linear reconstruction, was developed to align and combine multiple under-sampled dithered images from {\it HST} into a single image with improved resolution, reduction in correlated noise, and superior cosmic ray removal when compared to images combined using a lower quality shift-and-add method \citep{drizzlepac}.} {AstroDrizzle replaced MultiDrizzle in the {\it HST} data pipeline in June 2012 and is a significant improvement over the previous MultiDrizzle software as it directly utilizes the FITS headers for the instrument, exposure time, etc. instead of through user input.  AstroDrizzle also provides more freedom in {regard} to the parameters for the image combination, leading to faster, more compact, and target specific drizzled products  {\citep{adriz}}}. Using AstroDrizzle, we were able to adjust the parameters used in creating the median image, the shape of the kernel used in the final drizzled image, and the linear drop in pixel size when creating the final drizzled image, all of which allowed us to create products with sharper and smoother PSFs than previous MultiDrizzle or STScI pipeline products.

We processed each target in our sample in the same manner in order to best compare the final products. The 5 images in each filter were first registered using the {\tt tweakreg} task in DrizzlePac, which performed fine-alignment of the images via additional sources found using a {\tt daofind}-like algorithm. This fine-alignment was necessary to fully realize the high resolution of our observations to create accurate PSFs out of the drizzled products. After registering the images, they were combined through {\tt astrodrizzle}, which first drizzled each separate image, created a median image, and {split} the median image {back} into the separate exposures to convolve each exposure with the instrumental PSF and reconstruct it after the instrumental effects were removed. These reconstructed images were then corrected for cosmic ray contamination and finally drizzled together, with the final {\tt astrodrizzle} product scaled to $0\farcs03333/\rm{pixel}$. Lastly, we centered the target on a pixel to within $\pm0.01\,\rm{pix}$ by utilizing the {\tt astrodrizzle} output world coordinate system rotation matrix to transform the desired shift of the centroid of the star in pixel-space to a shift in RA/DEC-space. The drizzling and centering process was iterated as often as necessary to center the target on a pixel to the desired accuracy, which aided in constructing an accurate PSF.

Fig. \ref{fig:775driz} shows the final drizzled product in the F775W band for KIC 4139816, a typical single star from our sample. The {\it HST} pipeline product for this target showed a rough PSF near the center of the target, and further examination showed that the pipeline had incorrectly classified pixels in the saturated exposure. Manual adjustment of the data quality flags allowed us to correct the issue in our AstroDrizzled product, leading to a smoother and sharper PSF than the pipeline product.

\subsection{Kp$-${\it HST} Photometric Conversion}
\label{sec:kphst}

\begin{table}[t]
\begin{center}
\caption{Derived WFC3 photometry and $\rm{Kp}$ magnitudes from the \kepler Input Catalogue, used to derive Eq. \ref{eq:kphst}.}
\label{tab:phot}
\begin{tabular}{ccccc}
\hline \hline
KIC ID	&	Obs. Date	   &	$\rm{Kp}$	&	F555W	&	F775W	\\	\hline
2853029	&	2013-08-12 &	15.679		&	16.017	&	15.006	\\
4139816 	&	2013-04-12 &	15.954   		&	16.604 	&	15.141	\\
4813563	&	2012-11-12 &	14.254   		&	14.602 	&	13.510	\\
5358241	&	2013-02-04 &	15.386   		&	15.656 	&	14.902	\\
5942949	&	2012-10-29 &	15.699   		&	16.154 	&	14.990	\\	
6026438	&	2013-05-22 &	15.549   		&	16.075 	&	14.827	\\
6149553	&	2013-06-12 &	15.886   		&	17.004 	&	14.812	\\
6263593	&	2013-02-14 &	15.037		&	15.524	&	14.275	\\
6435936	&	2013-08-18 &	15.849   		&	16.846 	&	14.796	\\
7455287	&	2013-10-04 &	15.847   		&	16.720 	&	14.837	\\
8150320	&	2013-09-02 &	15.791   		&	16.303 	&	14.985	\\
8890150	&	2013-08-16 &	15.987   		&	16.853 	&	14.969	\\
8973129  	&	2013-07-07 &	15.056   		&	15.329 	&	14.455	\\
9838468  	&	2012-10-28 &	13.852   		&	14.108 	&	13.324	\\
10004738	&	2014-01-07 &	14.279		&	14.563	& 	13.704	\\
10118816	&	2012-10-27 &	15.233  		&	16.000 	&	14.226	\\	
10600955	&	2013-02-10 &	14.872   		&	15.135 	&	14.253	\\	
11305996	&	2013-03-31 &	14.807   		&	15.519 	&	13.850	\\
11497958	&	2013-04-06 &	15.921   		&	16.807 	&	14.805	\\
11768142	&	2013-07-31 &	15.931		&	17.056	&	14.895	\\
12256520	&	2013-07-28 &	14.477   		&	14.805 	&	13.957	\\
12470844	&	2013-03-19 &	15.339   		&	15.636 	&	14.695	\\
12557548	&	2013-02-06 &	15.692   		&	16.349 	&	14.936	\\	\hline 
\end{tabular}
\tablecomments{{\it HST} photometry is for blended stellar components in KIC 6263593, 11497958, and 11768142 systems. KIC 12557548 data are from \cite{croll13}. Observation Date is the same for all exposures of the same target.}
\end{center}
\end{table}

Converting the \kepler photometric system to the {\it HST} system served two purposes: the first to provide a check on the quality of our images and analysis, and the second to calculate the dilution of the transit depths due to additional stars in the \kepler photometric aperture. We calculated photometry from the AstroDrizzle products by summing the flux within a square aperture equivalent in area to a 2.0\arcsec radius aperture centered on the target. We then used the published encircled energy of 99\% relative to an infinite aperture along with published zero points\footnote{\url{www.stsci.edu/hst/wfc3/phot\_zp\_lbn} } to obtain F555W and F775W magnitudes for the targets. Errors on the magnitudes are estimated to be 0.03 in both filters.

\begin{figure}[t]
	\begin{center}
	\vspace{2pt}
	\includegraphics[width=0.45\textwidth]{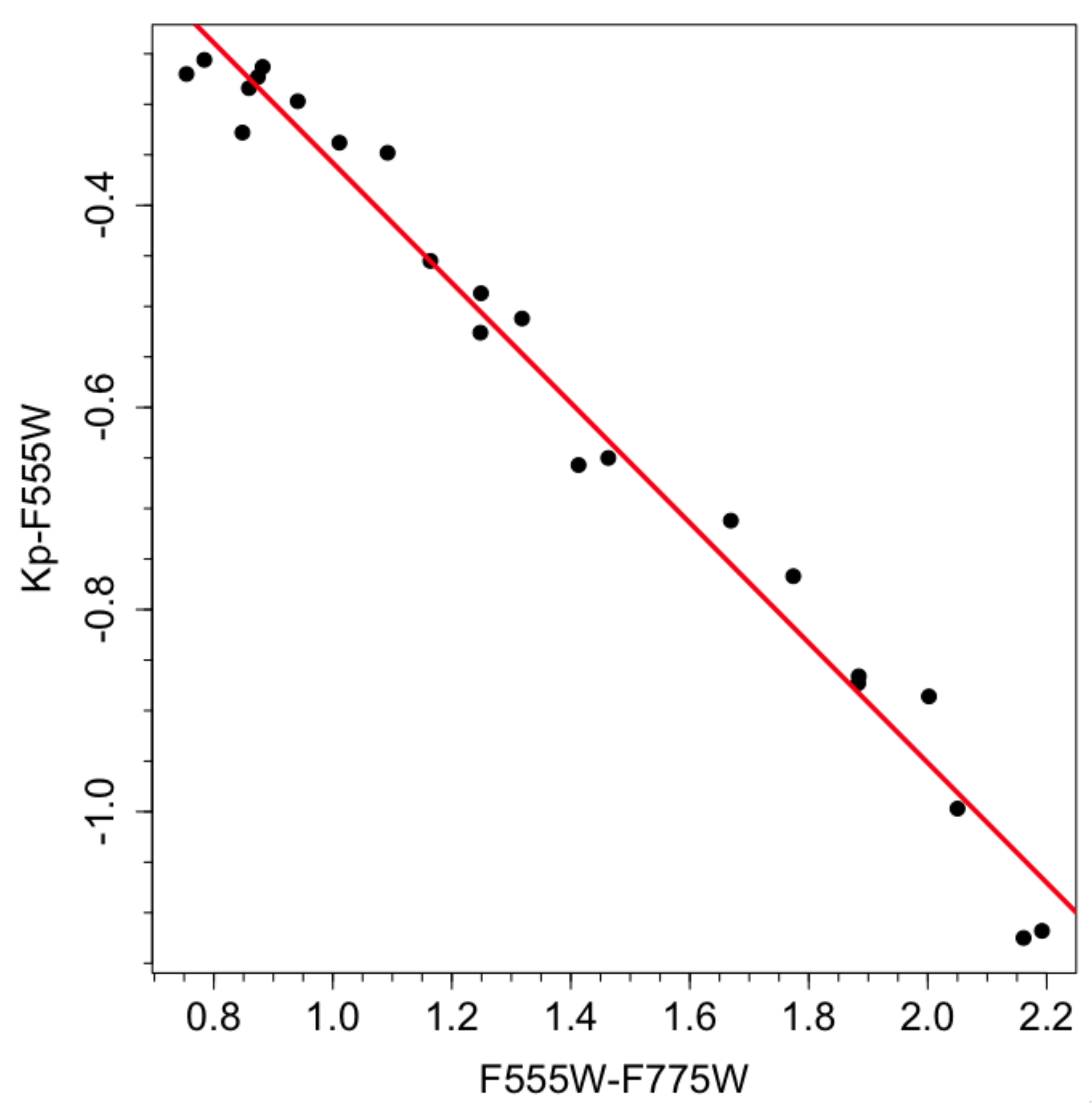}
	\caption{Plot of $\rm{Kp-F555W\, vs.\, F555W-F775W}$ (black points{, Table \ref{tab:phot}}) with the best fit linear model (Eq. \ref{eq:kphst}) plotted in red.  The tightness of the fit validates our echoice of a linear model to fit the conversion. The errors on fit and points are in the text.}     
	\label{fig:kphst}
	\end{center}
\end{figure}

We then compared the published values for $\rm{Kp}$ from the \kepler Input Catalogue to F555W and F775W for the 22 {observed targets }and one from \cite{croll13} that had identical observations (Table \ref{tab:phot}).  {Based on a plot of $\rm{Kp-F555W\, vs.\, F555W-F775W}$, we observed that the transformation between $\rm{Kp}$, F555W, and F775W would follow a linear relation.} Fitting of a linear model to the data produced the correlation shown in Fig. \ref{fig:kphst}, whose form follows
\begin{equation}
	\rm{Kp=	0.236+0.406\times F555W +0.594\times F775W}
	\label{eq:kphst}
\end{equation}


\noindent The fitted errors for this relation are 0.019 mag for the F555W and F775W coefficients and 0.027 mag for the intercept, with an RMS scatter about the fit of 0.042, showing that our simple linear modeling works well for this sample. The error on the derived Kp magnitude depends on the ${\rm F555W-F775W}$ color as

\begin{equation}
	\sigma_{\rm Kp} = \sqrt{0.019^2\,(\rm F555W-F775W)^2 + 0.027^2}
\label{eq:kperr}
\end{equation}

\noindent leading to slightly higher errors in Kp for redder targets in {\it HST}.
\section{Evaluation of {\it Kepler}-296, \triple, and \newbin\, Stellar Parameters}
\label{sec:multistar}

%

	\begin{figure}[t]
		\begin{center}
		\includegraphics[width=0.4\textwidth]{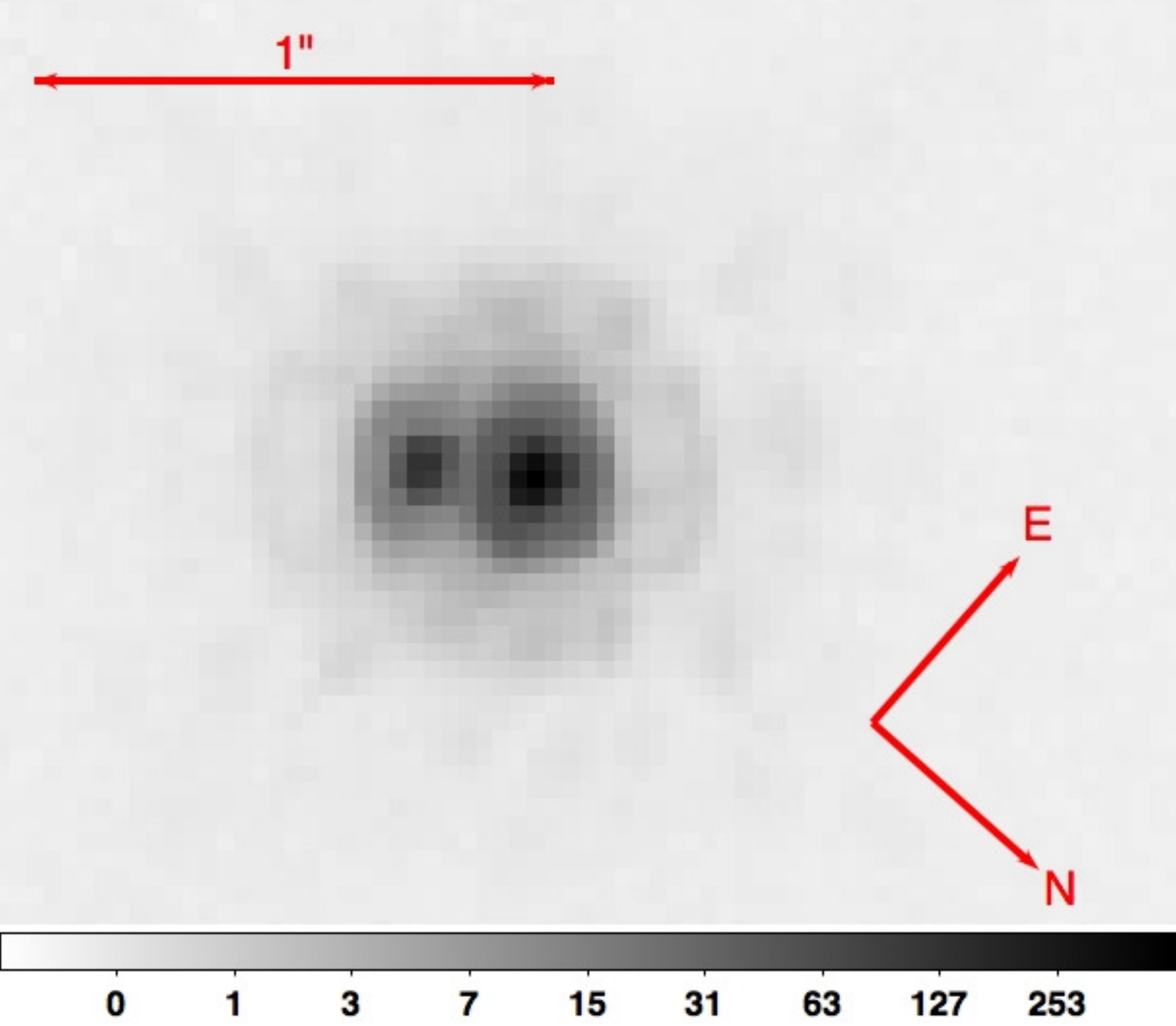}
		\caption{Drizzled image of \binary\, in the F775W filter showing a $1\farcs0$ scale bar and orientation. The fainter component, B, is to the left. Scale and units as in Fig. \ref{fig:775driz}. The FWHM of the PSF is $0\farcs1719$ for blended {system}.}     
		\label{fig:1422driz}
		\end{center}
	\end{figure}
	
	\begin{figure}[t]
		\begin{center}
		\includegraphics[width=0.4\textwidth]{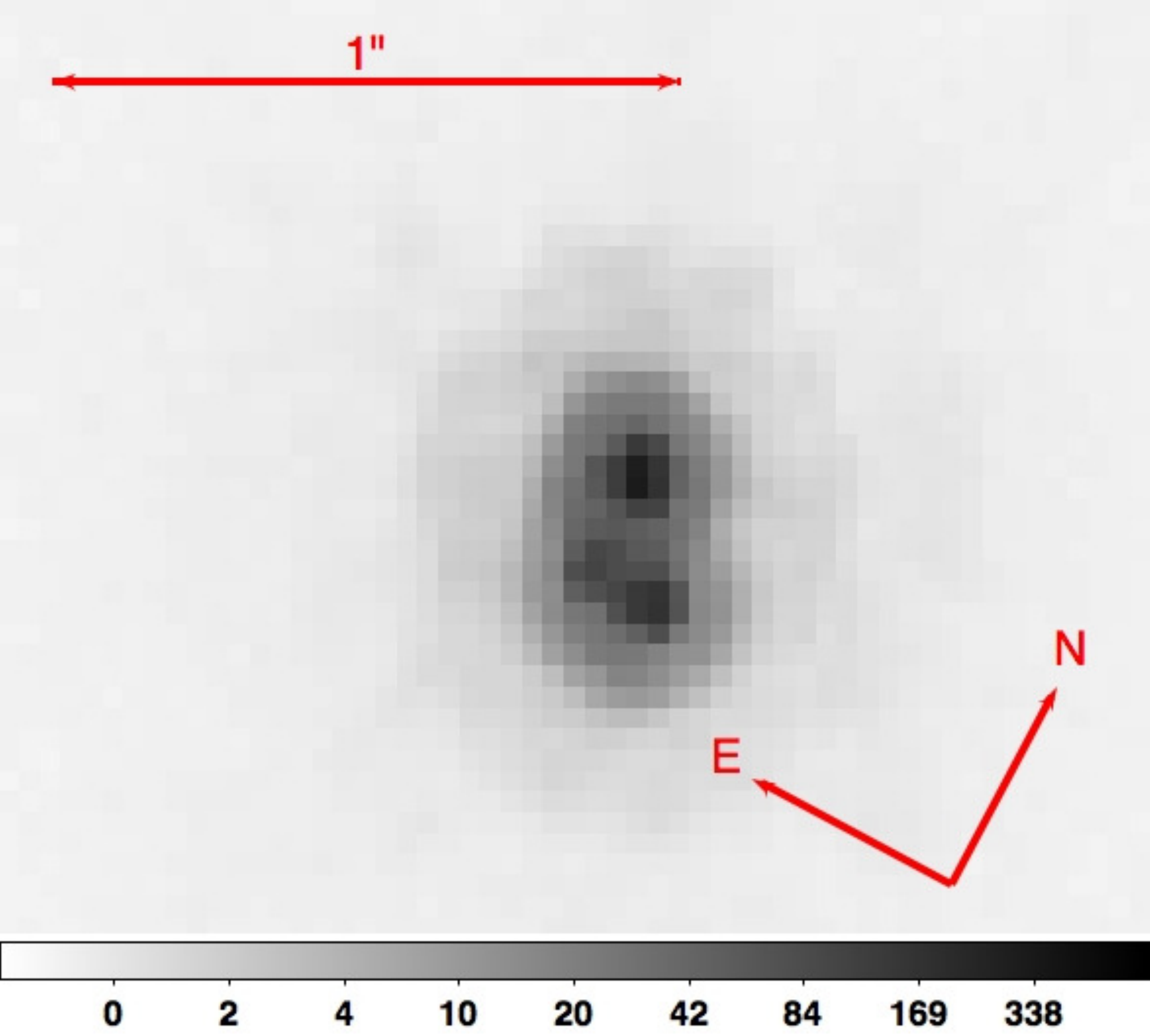}
		\caption{Drizzled image of \triple\, in the F775W filter showing a $1\farcs0$ scale bar and orientation. Component B is lowest in the image, with component C to the left. Scale and units as in Fig. \ref{fig:775driz}. The FWHM of the PSF is $0\farcs3870$ for blended {system}.}     
		\label{fig:2626driz}
		\end{center}
	\end{figure}

{Our program} observed three systems of particular interest: \binary, \triple, and \newbin. \binary\, was first published as a multiple planet system by \cite{borucki11} {and it has since been confirmed as a five planet system}. The stellar properties for this system were significantly updated by \cite{muir12}, \cite{dres13}, and \cite{mann13}, and as a result of these studies it was found that \binary\, contained at least three potentially habitable planets. However, \cite{liss14} showed using Keck AO and these {\it HST} images that \binary\, is actually a tight binary star system that appeared blended in the \kepler CCDs. \triple\, was first published in \cite{batalha13}, and examination by \citeauthor{dres13} showed that the single planet {candidate} in the system was potentially habitable, {though \cite{mann13} {disputed} this finding. Later Keck AO observations\footnote{\url{https://cfop.ipac.caltech.edu/edit\_obsnotes.php?id=2626; ``ciardi"}} {revealed \triple\, to be a tight triple star system, {and this realization} challenged all previous arguments about habitability}. It was noted in July 2013 on the {\it Kepler} Community Follow-up Observing Program (CFOP) {that Lick AO detected a secondary star in their image $0\farcs5$ away from \newbin\,\footnote{\url{https://cfop.ipac.caltech.edu/edit\_obsnotes.php?id=3049; ``hirsch"}}{(1 planet candidate)}, but no confirmation of association has been published to date.} The stellar multiplicity of each system has profound impacts on the habitability of their planets, which we re-evaluated in this study.

%

\begin{figure}[t]
	\begin{center}
	\includegraphics[width=0.4\textwidth]{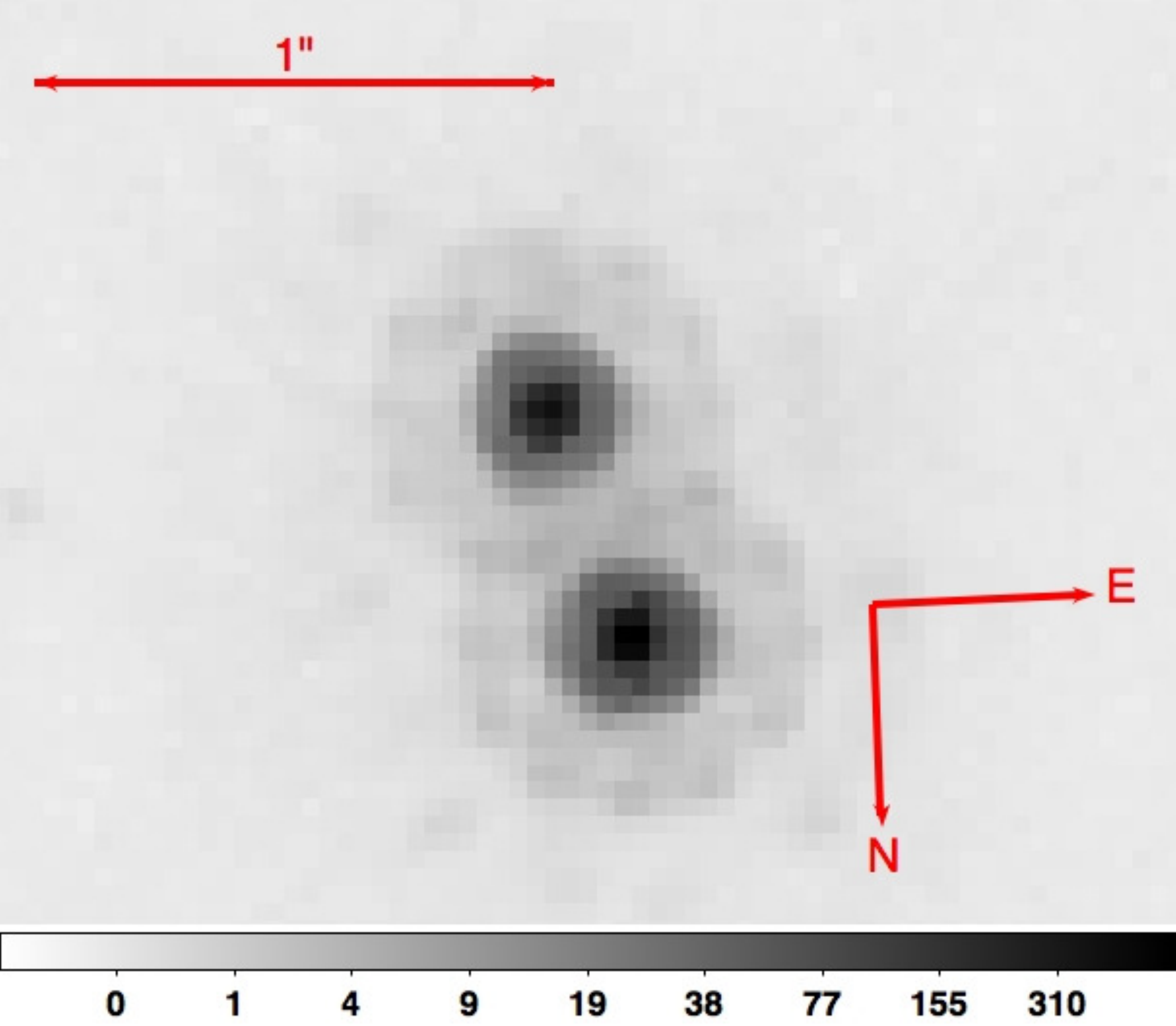}
	\caption{Drizzled image of \newbin\, in the F775W filter showing a $1\farcs0$ scale bar and orientation. The fainter component, B, is towards the top. Scale and units as in Fig. \ref{fig:775driz}. The FWHM of the PSF is 0.5563$\arcsec$ for blended {system}.}     
	\label{fig:3049driz}
	\end{center}
\end{figure}

Figures \ref{fig:1422driz}, \ref{fig:2626driz}, and \ref{fig:3049driz} show the AstroDrizzle combined images of \binary,\, \triple,\, and \newbin, respectively, and display the tight, apparent multiplicity of the systems.  We performed PSF fitting for each system as described in \cite{gs14} to {photometrically} separate the components in the {\it HST} filters. 

To ensure that the multiple components are not random superpositions of stars {at different distances}, we then attempted to fit the components of each system to a single isochrone to prove {that the systems' are  most likely bound and, therefore, that the stars are the same age (coeval). We then determined the probability that a random star in the field would produce a {false isochrone} match to the same precision while not being physically associated with the target star. This determines the probability of the isochrone fits for our target systems indicating bound systems over randomly superimposed stars on the CCD}. The PSF definition and the false association probability are outlined here and described in detail in \cite{gs14}.

\subsection{PSF Definition and Photometry Used}
\label{sec:psf}
{We adopted the global PSF solution of \cite{gs14} in each {\it HST} filter in order to separate the stellar components of each of the three systems. This global PSF was empirically generated from our observations of apparently single stars, and is a function of target color, {\it HST} focus (which changes by small amounts from thermal stresses), and sub-pixel centering of the target. We extracted the necessary parameters for the PSF from the drizzled image {of} each system of interest, and iteration of the PSF fitting returned the separation and orientations of the components of the systems and their fractional contributions in each {\it HST} bandpass. Lastly, combining the fractional contributions in the {\it HST} filters with the $\rm{Kp-{\it HST}}$ conversion in Eq. \ref{eq:kphst} returned the fractional contribution of light from each component in Kp, which is directly relevant to the planetary parameters inferred from the \kepler transit depth.} 

{Application of this algorithm for \binary\,shows that component} A contributes 80.9\% of the light in the \kepler bandpass, while component B contributes 19.1\% \citep{liss14}.  Estimated uncertainties for these percentages are 3\%. We found that component B is offset from the brighter component A by $0\farcs217\pm0\farcs004$ at a position angle of $217\fdg3\pm0\fdg8$ north through east.


\begin{table*}[t]
 \begin{center}
 \caption{Observed Photometry}
 \label{tab:obs}
 \begin{tabular*}{\textwidth} {@{\extracolsep{\fill} } c c c c c c c c}
  	\hline \hline
	\multicolumn{8}{c}{\binary\,Photometry} \\ \hline
	Star			&	F555W	&	F775W	&	{\it Ks}	&	Kp				&F555W-F775W	&	${i-J}$	&	F775W-{\it Ks}	\\ \hline
	A			&	16.997	&	15.040	&	\---		&	$16.076\pm0.045$	&1.957			&	\---		&	\---			\\
	B			&	18.874	&	16.396	&	\---		&	$17.641\pm0.053$	&2.478			&	\---		&	\---			\\
	${\rm A+B}$	&	16.820	&	14.766	&	\---		&	$15.845\pm0.047$	&2.053			&	1.807	&	\---			\\
	${\rm B-A}$	&	\---		&	1.356	&	\---		&	\---				&\---				&	\---		&	\---			\\ \hline	
	
	\multicolumn{8}{c}{\triple\,Photometry} \\ \hline
	Star			&	F555W	&	F775W	&	{\it Ks}	&	Kp				&F555W-F775W	&	${i-J}$	&	F775W-{\it Ks}	\\ \hline
	A			&	17.643	&	15.598	&	13.400	&	$16.669\pm0.047$	&2.045 			&	\---		&	2.198		\\
	B			&	18.406	&	16.107	&	13.838	&	$17.280\pm0.051$	&2.299			&	\---		&	2.269		\\
	C			&	19.289	&	16.900	&	14.520	&	$18.109\pm0.052$	&2.389			&	\---		&	2.380		\\
	A+B+C		&	17.057	&	14.886	&	12.634	&	$16.010\pm0.049$	&2.172			&	1.807	&	2.252		\\
	${\rm B-A}$	&	\---		&	0.509	&	0.438	&	\---				&\---				&	\---		&	\---			\\ 
	${\rm C-A}$	&	\---		&	1.302	&	1.120	&	\---				&\---				&	\---		&	\---			\\ \hline	
	
	\multicolumn{8}{c}{\newbin\,Photometry} \\ \hline	
	Star			&	F555W	&	F775W	&	{\it Ks}	&	Kp				&F555W-F775W	&	${i-J}$	&	F775W-{\it Ks}	\\ \hline
	A			&	16.004	&	14.806	&	\---		&	$15.537\pm0.035$	&1.198			&	\---		&	\---			\\
	B			&	16.646	&	15.284	&	\---		&	$16.080\pm0.037$	&1.362			&	\---		&	\---			\\
	${\rm A+B}$	&	15.526	&	14.266	&	\---		&	$15.022\pm0.036$	&1.259			&	1.209 	&	\---			\\
	${\rm B-A}$	&	\---		&	0.478	&	\---		&	\---				&\---				&	\---		&	\---			\\ \hline

   \end{tabular*}
 \tablecomments{Kp magnitudes and errors derived from Eq. \ref{eq:kphst} and \ref{eq:kperr}.}
 \end{center}
\end{table*}


{We used the same aforementioned global PSF and fitting algorithm} for \triple\,using the appropriate color, focus, and offset values. {We inspected the drizzled image minus the {PSF fit} for both F555W and F775W and found no evidence for yet further components in the \triple\, system.}  For \triple, component A contributes 54.5\% in the \kepler bandpass, component B contributes 31.0\%, and component C contributes 14.5\%. Estimated errors for these fractions are 6\%. We found that component B is separated from component A  by $0\farcs201\pm0\farcs008$ at a position angle of $212\fdg7\pm1\fdg6$, and component C is separated from component A by $0\farcs161\pm0\farcs008$ at $181\fdg6\pm1\fdg6$.

Fitting of the global PSF for \newbin\,using the corresponding color and focus values for this system showed that component A contributes 62.3\% in the \kepler bandpass and component B contributes 37.7\%, with estimated errors of 2\%. We found that component B is separated from component A by $0\farcs464\pm0\farcs004$ at a position angle of $196\fdg9\pm0\fdg8$. The estimated error for this system is lower than for either \binary\, or \triple\, as the components of the system are both brighter and more widely separated, and thus the PSF fitting was able to more distinctly separate the components.

In addition to the derived WFC3-based magnitudes and colors for the individual components of \binary, \triple, and \newbin, we also utilized the SDSS-based magnitudes \citep{sdss96} available in the \kepler Input Catalogue (KIC) \citep{kic11} as well as the 2MASS near-IR photometry available for the blended components. We found that the SDSS {\it g} and {\it r} band photometry was redundant for our late-type stars given our WFC3 photometry, and the SDSS {\it z} band was unreliable at the apparent magnitudes examined here \citep{kic11}. We therefore chose to include the blended photometry for the SDSS {\it i} band, adopting the transformation to standard SDSS photometry as detailed in \cite{pins12}. As 2MASS ${J-K}$ is relatively constant for a large span of early M dwarfs, we chose to utilize ${i-J}$ for the blended components in the fitting. Keck-AO data for \triple\, from NIRC-2 (Fig. \ref{fig:2626keck}) allowed PSF fitting to derive photometry for the individual components of that system in the {\it Ks} band which were used to replace the blended $i-J$ color in the isochrone fits. Our derived WFC3-based photometry, the blended ${i-J}$ colors, and the {\it Ks} band photometry for \triple\, used in the isochrone fitting are listed in Table \ref{tab:obs} for \binary,  \triple, and \newbin. We chose to use the $\Delta\rm{mag}$ in F775W {between components in} each system as the longer wavelength of that filter should be more reliable for our late-type stars than the F555W photometry.

	
		\begin{figure}[t]
		\begin{center}
		\includegraphics[width=0.4\textwidth]{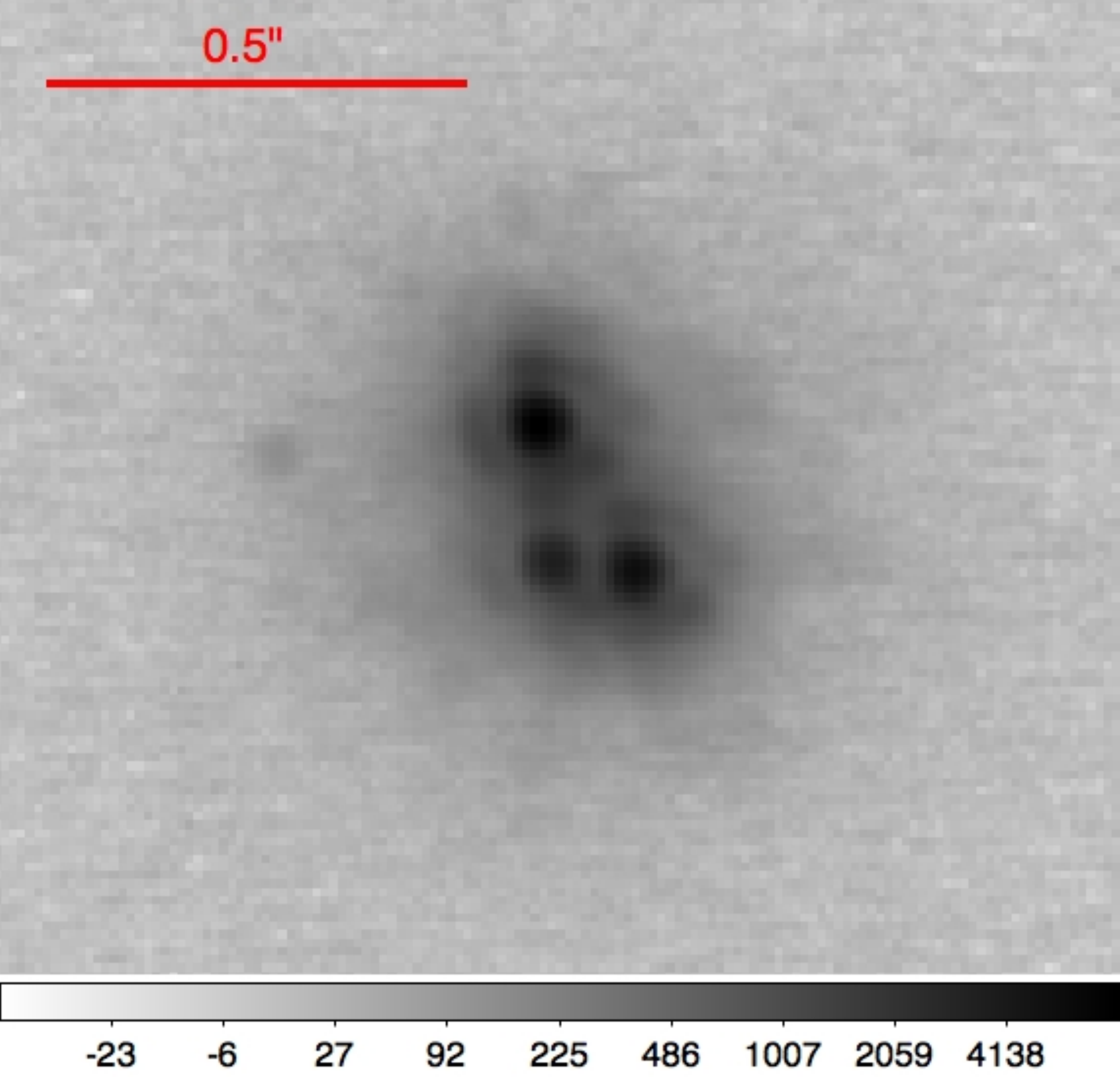}
		\caption{Keck K$^\prime$ image of \triple\, showing a $0\farcs5$ scale bar. Component A is highest in the image, with component B to the lower right and C to the lower left.}     
		\label{fig:2626keck}
		\end{center}
	\end{figure}

\subsection{Reddening Corrections}
\label{sec:red}

As we did not assume a distance (and therefore a reddening) value {\it a priori} for any of our systems, we allowed for adjustment of \ebv\,in order to find the best {isochrone} fit. We used the extinction laws for {\it J}, {\it i}, and {\it Ks} bands from \cite{pins12} which are

\begin{equation}
	\begin{matrix}
		\rm{A_{\it J}=0.282\times A_{\it V}} \\
		\rm{A_{\it i}=0.672\times A_{\it V}} \\
		\rm{A_{\it Ks}=0.117\times A_{\it V}}
	\end{matrix}
	\label{eq:aband}
\end{equation}

\noindent where $\rm{A_{\rm band}}$ is the extinction in the desired band and $\rm{A_{\it V}} = 3.1\times\ebv$ is the extinction in the {\it V} band. We calculated the extinction laws for F555W and F775W with the {\it HST} Exposure Time Calculator for WFC3/UVIS \footnote{\url{http://etc.stsci.edu/etc/input/wfc3uvis/imaging/} }, to be

\begin{equation}
	\begin{matrix}
		{\rm A_{F555W}=3.11}\times \ebv\\
		{\rm A_{F775W}=1.98}\times\ebv 
	\end{matrix}
	\label{eq:rband}
\end{equation}

\subsection{Fitting Using Victoria-Regina Isochrones}
\label{sec:isochrone}
Based on the derived WFC3 photometry for the components of \binary, \triple, and \newbin, we anticipated that \binary A {would match} the temperature of an early M dwarf, with \binary B a slightly later M dwarf  \citep{lepine13}. We also predicted \triple A to be a slightly later M dwarf than \binary A, \triple B between \binary A and \binary B, and \triple C slightly later than \binary B. We expected both \newbin A and \newbin B to be earlier types than \binary A, falling near late-K/early-M dwarfs \citep{kdwarf}. \cite{dres13} argue that the Dartmouth Stellar Evolution Database (DSED) \citep{dsed08} provides the most state-of-the-art representation of the evolution of M dwarfs and thus would provide reliable solutions for \binary, \triple, and \newbin.  \cite{feiden11} also demonstrated the reliability of the Dartmouth isochrones in fitting for late-type stars. 

We have found that the {DSED} isochrones systematically {underestimate} the temperatures, masses, and radii for M dwarfs when optical bandpasses are {relied upon} for the fitting. {The latest release of the DSED isochrones in 2012 utilizes the BT-Settl model atmosphere line lists and physics of \cite{btsettl}. The Dartmouth Stellar Evolution Program generated their synthetic photometry using the PHOENIX atmospheric code \citep{phoenixa, phoenixb} and {inputted} DSED boundary conditions from their isochrone grids. Thus, while the DSED isochrones did not use the exact model atmosphere grids released by \cite{btsettl}, the synthetic photometry included in the latest DSED release is still subject to the same strengths and weaknesses as the BT-Settl atmospheres.} Examination of Fig. 2 of \cite{btsettl} and Fig. 9 of \cite{mann13} shows that while the synthetic spectra for M dwarfs are remarkably accurate for infrared wavelengths, the molecular line lists for M dwarfs are incomplete in the optical and thus do not adequately represent the M dwarf spectral energy distribution in this wavelength range. These regions of the synthetic spectra are often masked out when attempting to use the BT-Settl atmospheric spectra to fit to observed M dwarf spectra. As BT-Settl appears to overestimate the SED of M dwarfs in the optical, inclusion of optical photometry when attempting to fit using BT-Settl photometry should always predict more optical flux than appears for a given stellar temperature, so would skew the fitting towards cooler temperatures. This is consistent with our comparison with \cite{dres13} (see \S\ref{sec:discussion} for more information). The synthetic photometry included in DSED predicts that below a certain temperature all M dwarfs have the same color in optical bandpasses, which does not match our full observational sample \citep{gs14}. The newest release of the Victoria-Regina (VR) Stellar Models {\citep{vdbiso, vdb, vdb2} } uses the MARCS model atmospheres that demonstrate increasingly red colors for decreasing stellar brightness, a much more accurate representation of observed M dwarfs in the solar neighborhood and our full target sample.

{The discrepancy in photometry tabulated in DSED and VR can be traced back to the differences between the latest PHOENIX \citep{btsettl} and MARCS \citep{vdb2} model atmosphere inputs and physics. To solve for the emergent intensity as a function of wavelength, MARCS uses a spherical 1D, local thermodynamic equilibrium (LTE) atmosphere while BT-Settl uses a spherically symmetric, LTE 2D solution with non-LTE physics for specific species. The most significant difference between these two atmospheric models are the molecular lines and opacities included in their calculations, as well as the inclusion of dust opacities, cloud formation, condensation, and sedimentation. BT-Settl includes all of the aforementioned advanced atmospheric calculations, while MARCS contains limited ionic and molecular opacities and no dust opacity or high-order atmospheric physics. As these details are most important for M dwarfs in the infrared, it logically follows that BT-Settl more accurately models stellar photometry in that range while the missing optical molecular bands in the PHOENIX models leads to inaccuracies in optical bandpasses \citep{btsettl, mann13}.}

Fig. \ref{fig:isocomp} shows solar, sub-solar, and super-solar metallicity, 5 Gyr isochrones from the VR and {DSED models} with stars from the RECONS project \citep{recons99, recons06, recons13, recons14} within 5 \pc\, of the Sun overplotted. {From this we can see that the stellar models are indistinguishable for stars with ${\rm F555W-F775W} $ colors bluer than $\sim1$.}  Stars with colors redder than 1 follow the  VR models more closely than the Dartmouth models. The deviation becomes greatest for colors redder than 2.5, where the RECONS data {show} a continual reddening of color with decrease in magnitude, which Dartmouth models do not show. Initial analysis using the Dartmouth isochrones yielded stellar temperatures that were significantly hotter than previous studies suggested \citep{dres13, muir12} and the lack of consistency with those calculations remained troubling until the limitations of Dartmouth models for cool stars in optical bandpasses were realized. We therefore used the synthetic photometry available for the VR isochrones for F555W, F775W, {\it i}, {\it J}, and {\it Ks} bands to perform our fitting.

{It has been noted in the past that stars in the solar neighborhood have a sub-solar average [Fe/H] metallicity \citep{hypatia}. Therefore, the RECONS stars should fall between the [Fe/H] = 0 and [Fe/H] = -0.5 isochrones in Fig. \ref{fig:isocomp}. The recently released Hypatia Catalog \citep{hypatia}, which compiles spectroscopic abundance data from 84 literature sources for 50 elements across 3058 stars within 150 \pc\,}{of the Sun, challenges this conclusion. After re-normalizing the raw spectroscopic data of their catalog stars to the same solar abundances, they find that the mean [Fe/H] for thin-disk stars in the solar neighborhood is +0.0643 and has a median value of +0.08. As the Hypatia Catalog indicates that solar neighborhood stars are actually slightly super-solar in metallicity, the location of the RECONS stars in relation to the VR isochrones in Fig. \ref{fig:isocomp} appears consistent.}

Using the data and codes provided by {\cite{vdbiso} and the interpolation methods described in Appendix A of} \cite{vdb2}, we generated ten 5 Gyr isochrones assuming a helium fraction of 0.27, [$\alpha$/Fe] = 0.0, and spanning the metallicity range ${\rm [Fe/H] = -0.5 \rightarrow +0.4}$ in steps of 0.1 dex. We then linearly interpolated the generated isochrones halfway between the given points and added calculations of ${\rm L/L_\odot}$ and  ${\rm R/R_\odot}$ from the quantities provided. The resulting isochrones contained synthetic photometry for F555W, F775W, {\it i}, {\it J}, and {\it Ks} bandpasses as well as fundamental stellar parameters. The final isochrones used spanned a range of $0.12\lesssim M_\star/M_\odot\lesssim1.2$.

The \kepler light curves for \binary, \triple, and \newbin\,all show low amplitude, long period variations ($\rm{\sim weeks}$) which are characteristic of older stars. As M-dwarfs evolve little over the course of their very long lives, we have adopted an age for all systems of 5 Gyr; adjustment of this age showed insignificant impact on the results. Assuming these are systems of late-type main sequence stars, we further restricted our isochrone fitting only to stars with $M_\star/\Msun\leq1.0$. Lastly, we required that the brightest component of each system be the most massive, with the dimmer component(s) being less massive. If the systems are truly bound then each component is at the same distance from us, meaning that the apparent magnitudes correlate with the effective temperatures and therefore with the mass.


\begin{figure}[t]
\begin{center}
\includegraphics[width=0.45\textwidth]{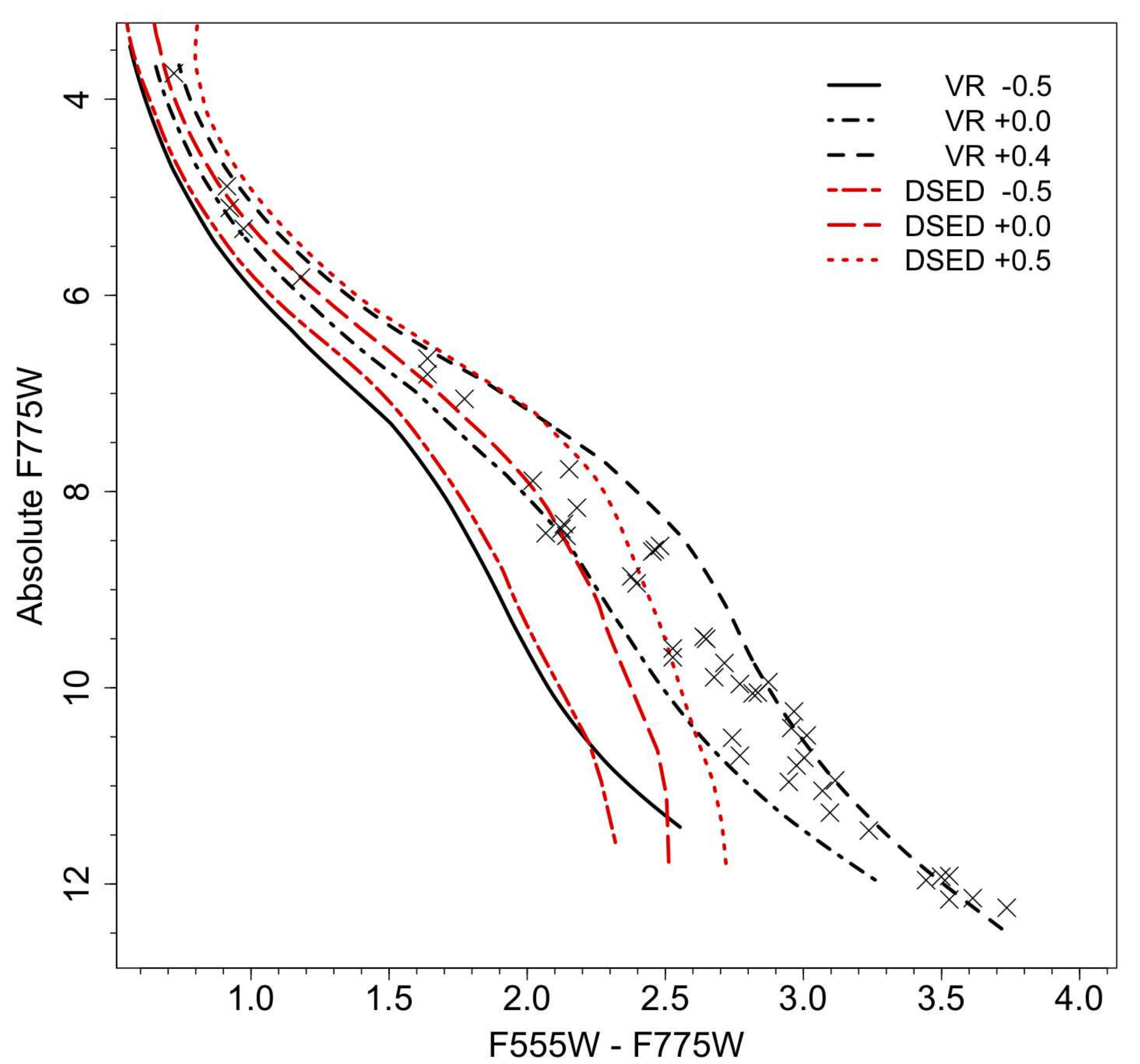}
\caption{Comparison of 5 Gyr isochrones from the Victoria-Regina Stellar Models (black) and the Dartmouth Stellar Evolution Database (red). Numbers in legend indicate the isochrone value of [Fe/H]. Crosses are stars within 5 \pc\, of the sun from the RECONS project with absolute photometry.}
\label{fig:isocomp}
\end{center}
\end{figure}

To fit both stellar components of \binary\, and \newbin\, {to an isochrone}, we performed a minimum-\chisq\, fitting between the observed and synthetic photometry described above. We chose to minimize the quadrature sum of the differences for the color of component A, the color of component B, the magnitude difference of B-A in F775W, and the blended ${i-J}$ color, given as
\begin{align}
	\label{eq:1422chisq}
	\chi^2_{\rm binary} & = (\Delta ({\rm F555W-F775W)_A} / \sigma_A)^2 \\  \nonumber
	&+ (\Delta ({\rm F555W-F775W)_B} / \sigma_{\rm B})^2 \\ \nonumber
	&+ (\Delta \,{\rm F775W_{B-A}} / \sigma_{\rm B-A})^2 \\ \nonumber
	&+ (\Delta (i-J)_{\rm A+B} / \sigma_{\rm A+B})^2	\nonumber
\end{align}

\noindent where $\rm \Delta(F555W-F775W) $ are the color differences between the observed colors and the tabulated values in the synthetic VR isochrones, $\rm \Delta F775W_{\rm B-A}$ is the observed difference in magnitude between components B and A in the F775W band minus the same quantity from the isochrones, and $\Delta (i-J)_{\rm A+B}$ is the $i - J$ color for the observed blended A+B photometry minus the blended isochrone values for A+B. The $\sigma$ values {represent the uncertainties in the measured photometry and }were set to 0.03 mag for \binary\, and 0.02 mag for \newbin\, for colors within the same photometric system, and 0.08 for cross-system colors (i.e. for $\it i - J$).

For the three components of \triple, we performed a similar minimum-\chisq\, fitting, including {\it Ks} band photometry in place of $i-J$ and adding appropriate terms for component C, given as
\begin{align}
	\label{eq:2626chisq}
	\chi^2_{\rm triple} & = (\Delta {\rm (F555W-F775W)_A} / \sigma_{\rm A})^2 \\  \nonumber
	&+ (\Delta {\rm (F555W-F775W)_B} / \sigma_{\rm B})^2 \\ \nonumber\displaybreak[1]
	&+ (\Delta {\rm (F555W-F775W)_C} / \sigma_{\rm C})^2 \\ \nonumber\displaybreak[1]
	&+(\Delta {\rm (F775W-{\it Ks})_A} / \sigma_{\rm A})^2 \\  \nonumber\displaybreak[1]
	&+ (\Delta {\rm (F775W-{\it Ks})_B} / \sigma_{\rm B})^2 \\ \nonumber\displaybreak[1]
	&+ (\Delta {\rm (F775W-{\it Ks})_C} / \sigma_{\rm C})^2 \\ \nonumber\displaybreak[1]
	&+ (\Delta \,{\rm F775W_{B-A}} / \sigma_{\rm B-A})^2 \\ \nonumber\displaybreak[1]
	&+ (\Delta \,{\rm F775W_{C-A}} / \sigma_{\rm C-A})^2 \\ \nonumber\displaybreak[1]
	&+ (\Delta \,{\rm {\it Ks}_{B-A}} / \sigma_{\rm B-A})^2 \\ \nonumber\displaybreak[1]
	&+ (\Delta \,{\rm {\it Ks}_{C-A}} / \sigma_{\rm C-A})^2 \nonumber\displaybreak[1]
\end{align}

\noindent Terms in Eq. \ref{eq:2626chisq} are the same as Eq. \ref{eq:1422chisq}, with the addition of $\rm \Delta(F555W-F775W) $ for the C component, $\rm \Delta F775W_{C-A}$ for the observed difference in magnitude between components C and A in the F775W band minus the same quantity from the isochrones, and similar quantities for F775W-{\it Ks} colors and $\Delta${\it Ks} magnitudes of all components. The $\sigma$ values in Eq. \ref{eq:2626chisq} were set to 0.05 mag {for} all terms except any involving component C, which were set to 0.08. The $\sigma$'s were increased to account for the larger uncertainty in the PSF fitting and thus the contributions of each component to the total magnitude. When fitting the observed photometry to the isochrones, we used the reduced \chisq\, metrics, where \chisq$_{\rm binary}$ was reduced by a factor of $(1-{\rm d.o.f.})=3$ and \chisq$_{\rm triple}$ was reduced by a factor of $(1-{\rm d.o.f.})=9$.

{In the fitting} of \binary\, and \newbin, for each {primary mass value ($M_A$), the secondary mass value ($M_B$) }that produced the minimum \chisq\, as per Eq. \ref{eq:1422chisq} was selected, assuming {$M_B < M_A$}. The overall best {isochrone match} was the combination of A and B masses that produced the global minimum \chisq$_{\rm binary}$. This two-level fitting was performed for the three binary permutations of components of \triple\, as well, to determine that each binary permutation of the system (A-B, A-C, and B-C) could also be coeval, to ensure that the photometry was producing consistent results between combinations of components, and to provide initial values for the masses of each component in the triple-star fitting. To perform the three-component fitting, we took the initial estimates for the masses of each component and searched a range of surrounding masses for the best fit, with the size of the range dependent on the reliability of the photometry for that component. For each mass in the range of component A, Eq. \ref{eq:2626chisq} was minimized for every combination of B and C masses. The overall combination of A, B, and C, that produced the global minimum of \chisq$_{\rm triple}$ was adopted as the best fit. 

{In order to test the systematic uncertainties in using the VR isochrones to determine the stellar mass, radius, and bolometric luminosity of our three target systems, we applied an offset to the solar metallicity VR model in order to match the RECONS stars in Fig. \ref{fig:isocomp}. We then fit the isochrones with the offset to \binary\, according to the method described above to test how the slight offset in metallicity affects the determination of the stellar parameters. We first fit the solar metallicity isochrone to the \binary\, photometry as is, then did the same by applying a shift in F555W-F775W color to match RECONS colors, and finally by applying a shift in F775W magnitude to match the RECONS magnitudes. This yielded two measurements of the systematic uncertainty when fitting for mass, radius, and luminosity. We find that the VR models required a shift of $\Delta{\rm F775W}=-0.5$ or $\Delta{\rm(F555W-F775W)}=+0.2$ in order to best match the RECONS sample.{We note that} the chosen shift in color matches the colors of the cooler stars in the sample while being slightly too red to properly match the hotter stars. The shift in magnitude did not affect the fit at all since the search range to match the magnitudes of the \binary\, components was larger than the model shift and so the fitting algorithm still selected the minimum $\chi^2$ fit. To calculate the systematic uncertainty of our isochrone fitting we averaged the {differences} between the best fit stellar parameters and the color-shifted best fit stellar parameters for the primary and secondary stars in \binary. {We find that  $\Delta\rm M=-0.081 M_\odot$, $\Delta \rm R=-0.071R_\odot$, $\Delta\rm L=-0.014L_\odot$, and $\Delta\rm T_{eff}=-154.55{\rm K}$.} From this we conclude that the systematic uncertainties when fitting for stellar mass, radius, and luminosity are small, but not insignificant, contributions to the total error budget.}

Lacking spectroscopic determinations for metallicity for \binary, \triple, or \newbin, we fit each system to isochrones of each metallicity in our range at \ebv = 0 to find the best fitting metallicity, and then increased the reddening to determine whether {that would provide a better fit}. In all cases, \ebv=0 provided the best fits. Table \ref{tab:metal} provides the minimum \chisq\, for each system at each metallicity for \ebv=0. \binary\, and \triple\, both show a clear best fit for [Fe/H] = +0.3 and +0.1, respectively. While \newbin\, has a best fit for [Fe/H] $= -0.4$, all metallicities tested show approximately the same goodness of fit, suggesting the independence of the {goodness-of-fit} with {regard} to metallicity for that system {and an even weaker assertion about the true metallicity of \newbin.} For the evaluation of planetary habitability, stellar parameters from the best fit metallicity (highlighted in bold in Table \ref{tab:metal}) were chosen. As the best fit \chisq\,for \binary\, is significantly below 1, we are likely overestimating our errors for that system.

\begin{table}[t]
 \begin{center}
 \caption{Values of the min \chisq\, for changing values of metallicity for \binary, \triple, and \newbin.}
 \label{tab:metal}
  \begin{tabular}{cccc}
  	\hline \hline
	[Fe/H]&	\binary 	&	\triple	&	\newbin		 \\ \hline
  	-0.5	&	3.187	&	1.610	&	0.936	\\ 
	-0.4	&	3.187	&	1.491	&	{\bf 0.908}	\\ 
	-0.3	&	6.227	&	1.313	&	1.056	\\ 
	-0.2	&	7.531	&	1.191	&	1.179	\\
	-0.1	&	8.365	&	1.139	&	1.086	\\
	0.0	&	6.246	&	0.941	&	0.943	\\
	+0.1	&	3.207	&	{\bf 0.860}	&	1.049	\\
	+0.2	&	0.704	&	1.258	&	1.073	\\
	+0.3	&	{\bf 0.218}	&	2.123	&	1.039	\\
	+0.4	&	1.568	&	3.987	&	1.041	\\	\hline
   \end{tabular}
 \end{center}
\end{table}

\subsection{False Association {Odds}}
\label{sec:fp}
In addition to showing that the suspected companion stars for \binary, \triple, and \newbin\, are coeval, we performed a Bayesian-like {odds ratio} analysis on the three systems to determine the probability that the isochrone fitting described in \S\ref{sec:isochrone} could have produced a good match {for all components }without the stars being physically associated \citep{gs14}. For the components of \binary, the {odds} ratio associated:random was 4101.6:1; for \triple, the ratio was 2832.9:1 for the primary and secondary companions and 928.1:1 for the primary and tertiary companions; for \newbin\ the ratio was 1923.7:1. {From this we conclude that isochrone fitting utilizing the photometry of these three cases would be very unlikely to produce a good fit if the stars were random superpositions and not truly associated.}

\subsection{{\it Kepler}-296 Best-fit Stellar Parameters}
\label{sec:1422fit}

\begin{figure*}[t]
	\begin{center}
	\includegraphics[width=0.7\textwidth, height=3.0in]{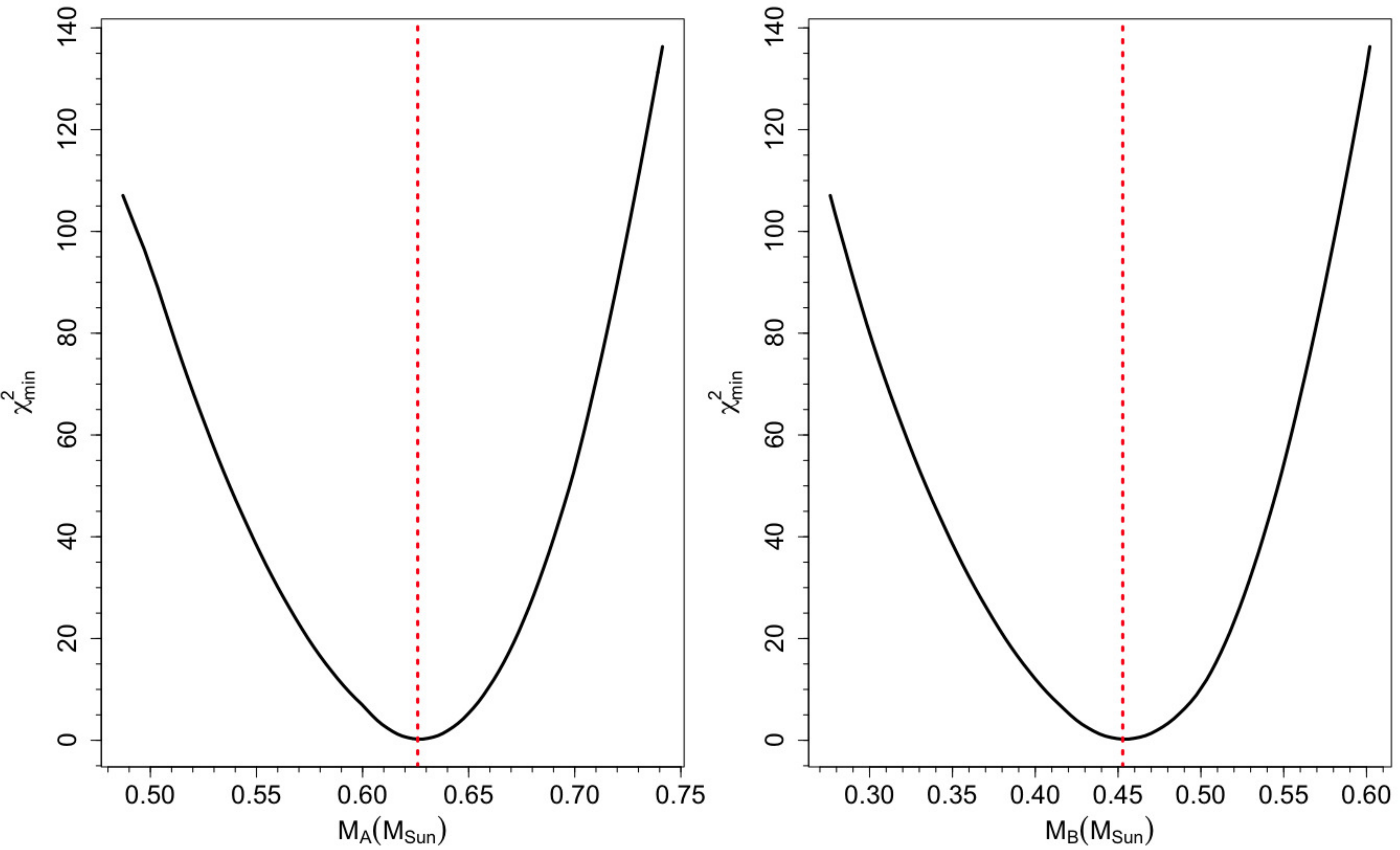}
	\caption{Left: variation of \chisq\, from Eq. \ref{eq:1422chisq} for $M_\star/\Msun$ for component A of \binary . Right: same as left panel, for component B of \binary . Black curve shows the variation of \chisq , red dashed line shows mass of components for the minimum \chisq .}     
	\label{fig:1422chisq}
	\end{center}
\end{figure*}

\begin{table}[t]
 \begin{center}
 \caption{Best fit stellar parameters for the components of {\it Kepler}-296}
 \label{tab:1422param}
  \begin{tabular}{c c c}
  	\hline \hline
	Parameter			&	\binary A			&	\binary B		\\ \hline
	$M_\star/\Msun$		&	$\binma\pm\binmaerr$	&	$\binmb\pm\binmberr$	\\
	$\teff\, [K]$				&	$\binta\pm\bintaerr$	&	$\bintb\pm\bintberr$	\\
	$R_\star/\Rsun$		&	$\binra\pm\binraerr$	&	$\binrb\pm\binrberr$	\\
	Distance [pc]			&	$359$			&	$358$				\\
	F555W				& 	$ 9.218$			& 	$11.111$			\\
	F775W				&	$7.266 $			&	$8.621$			\\
	$\rm{F555W - F775W}$	&	$1.952$			&	$2.490$			\\ 
	$\rm{F775W_{B-A}}$	&	\multicolumn{2}{c}{$1.356$}				\\\hline
   \end{tabular}
 \tablecomments{Tabulated values were calculated for $\ebv=0.00$, [Fe/H] = +0.3, age = 5 Gyr, and were matched to the observed values in Table \ref{tab:obs}. \chisq$_{min}$ = \binchisq.}
 \end{center}
\end{table}

Using the procedures described in \S\ref{sec:isochrone} and \S\ref{sec:red} we found that the best fit for the stellar components of \binary \,occurred for [Fe/H] = +0.3, with $M_A/\Msun=\binma\pm\binmaerr$ and $M_B/\Msun=\binmb\pm\binmberr$. The tabulated temperatures that correspond to these masses in the VR isochrones are $T_A=\binta\pm\bintaerr \rm{\,K}$ and $T_B=\bintb\pm\bintberr\rm{\,K}$. These roughly correspond to spectral types M0.0V and M3.0V, respectively, based on the \cite{lepine13} spectroscopic catalogue of the brightest K and M dwarfs in the northern sky, which provided ranges and average temperature for each spectral subtype. The stellar radii are $R_A/\Rsun=\binra\pm\binraerr$ and $R_B/\Rsun=\binrb\pm\binrberr$, as calculated from the tabulated values of $\teff$ and stellar luminosity from the isochrones. {Errors on all of these values are $\delta_{\rm X} = \sqrt{1\sigma_{\rm iso}^2 + \Delta(\rm X)^2}$, where $1\sigma_{\rm iso}$ are the 1$\sigma$ errors above the minimum reduced \chisq\, value of \binchisq\, from the isochrone fitting and $\Delta(\rm X)$ are the systematic uncertainties in the isochrone fitting as described in \S\ref{sec:isochrone}.} Fig. \ref{fig:1422chisq} shows the variation of \chisq\, (calculated as in Eq. \ref{eq:1422chisq}) with the best-fit masses of the primary and secondary component of \binary\, indicated. The $1 \sigma_{\rm iso}$ errors were calculated by finding the two points along the \chisq\, curves in Fig. \ref{fig:1422chisq} that corresponded to values of \chisq$_{min} + 1.57$, accounting for 4 degrees of freedom in the fit \citep{numrec}. The optimal stellar parameters and their errors are tabulated in Table \ref{tab:1422param}. 

We calculated the distance to \binary\, by applying the distance modulus formula to the observed and absolute magnitudes of each component in each {\it HST} filter then averaging the four estimates. The absolute magnitudes from the isochrone match combined with the apparent magnitudes from our {\it HST} imaging implies a distance to \binary\, of $\bind\pm\binderr\,\pc$. At this distance, the empirically measured separation of $0\farcs217\pm0\farcs004$ translates to a physical separation of $\bina\pm\binaerr\,\au$ and an orbital period of $\binp\pm\binperr$ years. The true values of both the separation and period are likely larger due to projection effects foreshortening the true separation and orbital period.

\subsection{KOI-2626 Best-fit Stellar Parameters}
\label{sec:2626fit}

\begin{table}[t]
 \begin{center}
 \caption{Best fit stellar parameters for the components of \triple}
 \label{tab:2626param}	
  \begin{tabular}{c c c c}
 	\hline \hline
	Parameter			&	\triple A			&	\triple B			&	\triple C			\\ \hline
	$M_\star/\Msun$		&	$\tripma\pm\tripmaerr$	&	$\tripmb\pm\tripmberr$	&	$\tripmc\pm\tripmcerr$	\\
	$\teff\, [K]$			&	$\tripta\pm\triptaerr$	&	$\triptb\pm\triptberr$	&	$\triptc\pm\triptcerr$	\\
	$R_\star/\Rsun$		&	$\tripra\pm\tripraerr$	&	$\triprb\pm\triprberr$	&	$\triprc\pm\triprcerr$	\\
	Distance [pc]			&	337				&	342				&	333				\\
	F555W 				&	$10.007$ 	 		&	$10.697$ 	 		&	$11.690$			\\
	F775W				&	$7.953$	 		&	$8.472$ 			&	$9.274$			\\
	{\it Ks}				&	$5.732$	 		&	$6.151$ 			&	$6.839$			\\
	$\rm{F555W - F775W}$	&	$2.054$			&	$2.225$			&	$2.416$			\\
	$\rm{F775W - {\it Ks}}$	&	$2.221$			&	$2.321$			&	$2.435$			\\
	$\rm{F775W_{B-A}}$	&	\multicolumn{2}{c}{$0.518$}				&					\\
	$\rm{F775W_{C-A}}$ 	& 	\multicolumn{2}{c}{$1.321$}				&					\\
 	$\rm{{\it Ks}_{\,B-A}}$	&	\multicolumn{2}{c}{$0.420$}				&					\\
	$\rm{{\it Ks}_{\,C-A}}$ 	& 	\multicolumn{2}{c}{$1.107$}				&					\\	\hline
   \end{tabular} 
  \tablecomments{Tabulated values were calculated for $\ebv=0.00$, [Fe/H] = +0.1, age = 5 Gyr, and were matched to the observed values in Table \ref{tab:obs}. \chisq$_{min}$ = \tripchisq.}
     \vspace{-0.2cm}
 \end{center}
\end{table}

The best fit for \triple\, occurred for [Fe/H] = +0.1, with $M_A/\Msun=\tripma\pm\tripmaerr$, $M_B/\Msun=\tripmb\pm\tripmberr$, and $M_C/\Msun=\tripmc\pm\tripmcerr$. The tabulated temperatures that correspond to these masses in the VR isochrones are $T_A=\tripta\pm\triptaerr \rm{\,K}$,  $T_B=\triptb\pm\triptberr\rm{\,K}$, and $T_C=\triptc\pm\triptcerr\rm{\,K}$. These temperatures translate roughly to M1.0V, M2.0V, and M2.5V, respectively based on \cite{lepine13}. The stellar radii are $R_A/\Rsun=\tripra\pm\tripraerr$, $R_B/\Rsun=\triprb\pm\triprberr$, and $R_C/\Rsun=\triprc\pm\triprcerr$ as calculated from the tabulated values of $\teff$ and stellar luminosity from the isochrones. These parameters are tabulated in Table \ref{tab:2626param}. Curves showing the variation of \chisq\, (calculated as in Eq. \ref{eq:2626chisq}) as a function of stellar mass similar to Fig. \ref{fig:1422chisq} were created and used to determine the best fit {and $1\sigma_{\rm iso}$ points. The listed errors are calculated as in \S\ref{sec:1422fit} with $1\sigma_{\rm iso} = $\chisq$_{min} + 1.28$ above the minimum \chisq\, value of \tripchisq, accounting for the 10 degrees of freedom in the fitting \citep{numrec}.}

The absolute magnitudes from the isochrone match combined with the apparent magnitudes from our {\it HST} imaging implies a distance to \triple\, of $\tripd\pm\tripderr\rm{\,pc}$. At this distance, the empirically measured separation of $0\farcs203$ between components A and B translates to a physical separation of $\tripab\pm\tripaberr\,\au$ and for the measured separation of components A and C of $0\farcs161$ we calculated a physical separation of $\tripac\pm\tripacerr\,\au$. Again, the real values are likely larger due to projection effects.


\subsection{KOI-3049 Best-fit Stellar Parameters}
\label{sec:3049fit}

\begin{table}[t]
 \begin{center}
 \caption{Best fit stellar parameters for the components of \newbin}
 \label{tab:3049param}
\begin{tabular}{c c c}
  	\hline \hline
	Parameter			&	\newbin A			&	\newbin	B		\\ \hline
	$M_\star/\Msun$		&	$\newbinma\pm\newbinmaerr$	&	$\newbinmb\pm\newbinmberr$	\\
	$\teff\, [K]$				&	$\newbinta\pm\newbintaerr$	&	$\newbintb\pm\newbintberr$	\\
	$R_\star/\Rsun$		&	$\newbinra\pm\newbinraerr$	&	$\newbinrb\pm\newbinrberr$	\\
	Distance [pc]			&	485				&	484			\\
	F555W				& 	$7.567$			& 	$8.222$			\\
	F775W				&	$6.381$			&	 $6.858$			\\
	$\rm{F555W - F775W}$	&	$1.186$			&	$1.364$			\\ 
	$\rm{F775W_{B-A}}$	&	\multicolumn{2}{c}{$0.478$}				\\\hline
   \end{tabular}
 \tablecomments{Tabulated values were calculated for $\ebv=0$, [Fe/H] = -0.4, age = 5 Gyr, and were matched to the observed values in Table \ref{tab:obs}. \chisq$_{min}$ = \newbinchisq.}
 \end{center}
\end{table}


The best fit for the components of \newbin \,occurred for ${\rm [Fe/H]} = -0.4$. We find that $M_A/\Msun=\newbinma\pm\newbinmaerr$ and $M_B/\Msun=\newbinmb\pm\newbinmberr$. The tabulated temperatures that correspond to these masses in the VR isochrones are $T_A=\newbinta\pm\newbintaerr \rm{\,K}$ and $T_B=\newbintb\pm\newbintberr\rm{\,K}$. These effective temperatures match approximately to K4.0V and K5.5V, respectively, based on the spectral types tabulated in \cite{kdwarf}, as the temperatures are outside the range provided by \cite{lepine13}. We find the stellar radii to be $R_A/\Rsun=\newbinra\pm\newbinraerr$ and $R_B/\Rsun=\newbinrb\pm\newbinrberr$. The optimal stellar parameters and their errors are tabulated in Table \ref{tab:3049param}. Curves showing the variation of \chisq\, (calculated as in Eq. \ref{eq:1422chisq}) as a function of stellar mass  similar to Fig. \ref{fig:1422chisq} were created and used to determine the best fit and $1\sigma$ points. {The listed errors are determined as in \S\ref{sec:1422fit}, with $1\sigma_{\rm iso}$ calculated using the minimum \chisq\, value of \newbinchisq. }

The absolute magnitudes from the isochrone match combined with the apparent magnitudes from our {\it HST} imaging implies a distance to \newbin\, of $\newbind\pm\newbinderr\,\pc$. At this distance, the empirically measured separation of $0\farcs464\pm0\farcs004$ translates to a physical separation of $\newbina\pm\newbinaerr\,\au$ and an orbital period of $\newbinp\pm\newbinperr$ years. Again, the true values are likely larger due to projection effects.

\subsection{Isochrone Fit Discussion}


\begin{figure}[t]
	\begin{center}
	\includegraphics[width=0.45\textwidth]{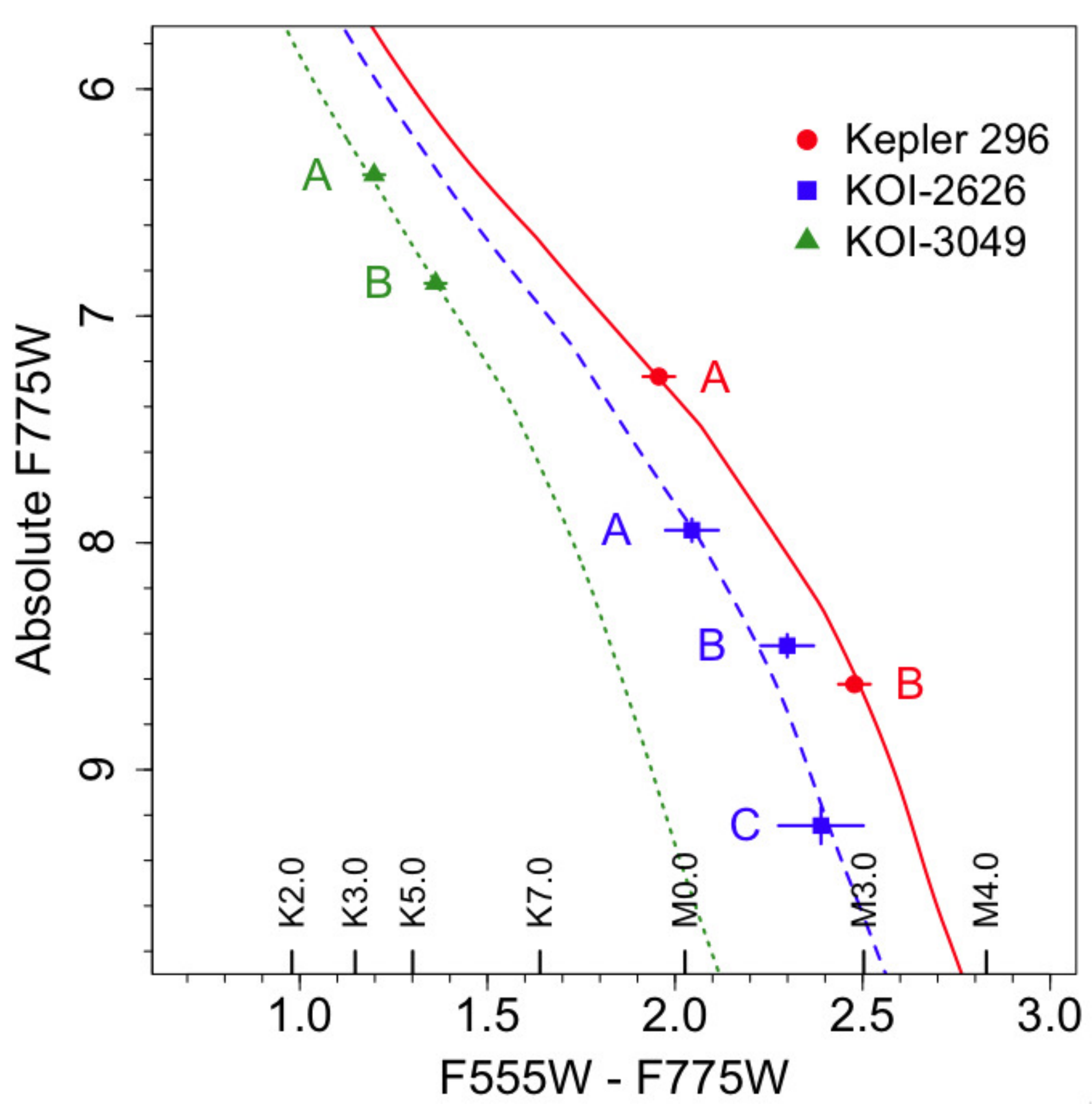}
	\caption{Absolute photometry of stellar components of \binary,\, \triple,\, and \newbin\,plotted over their respective best fit 5 Gyr isochrones. \binary\, components are in red circles plotted over an [Fe/H] = +0.3 isochrone (red solid line), \triple\, components are in blue squares plotted over an [Fe/H] = +0.1 isochrone (blue dashed), \newbin\, components are in green triangles plotted over an [Fe/H] = -0.4 isochrone (green dotted). Error bars are $1\,\sigma$. Spectral types are from \cite{lepine13} for types later than K6.0 and from \cite{kdwarf} for types earlier than K6.0.}    
	\label{fig:isochrone}
	\end{center}
\end{figure}


To compare the best-fit stellar properties of \binary, \triple, and \newbin\, we plotted each component atop their respective best fit isochrones in Fig. \ref{fig:isochrone}. The observed photometry tabulated in Table \ref{tab:obs} was converted to absolute photometry using the distances derived from the respective isochrone fits. From Fig. \ref{fig:isochrone} we note that our initial guesses at the relative magnitudes of the components of all three systems were correct, and that \binary, and \newbin\, are very likely bound binary systems based on their close fits to the VR isochrones. The only star that falls somewhat off of the isochrone is \triple\,B, which appears to be slightly redder than the isochrone fit would suggest. However as \triple\,B still fits the isochrone within its $1\sigma$ error on color, we still report with high confidence that \triple\, is a bound triple star system. 

\section{Planetary Habitability}
\label{sec:planets}

The multiplicity of \binary, \triple, and \newbin\, have interesting implications on the habitability of the planets in each system. \cite{dres13} determined that the planets \binary\,{\it d} (the third planet in the system) and \triple.01 (the only detected planet {candidate} in the system) were habitable, given the systems' previously assumed single-star properties. \cite{mann13} re-evaluated the temperatures of these stars using stellar temperatures derived from mid-resolution spectra and found that those two planets were actually interior to their respective Habitable Zones. However, neither of those studies accounted for the multiplicity of those systems, and thus their HZ analyses are {inaccurate for these targets}. Knowing now that \binary, \triple, and \newbin\, are multiple-star systems, we recalculated the planetary parameters of all detected planets around each potential stellar host using the best-fit stellar parameters in order to re-evaluate the planetary habitability. 


Circumbinary and circum-triple {planetary} orbits were not tested for habitability, as the wide physical separations of the systems coupled with the short transit periods preclude {planetary} orbits around multiple stars. Our projected separations of the stellar components of \binary, \triple, and \newbin\, indicate that they are either close or moderately separated systems, but as we cannot correct for projection effects the systems could {be more widely separated}. While circum-primary orbits reduce the likelihood of the additional stellar component(s) interacting catastrophically with the planetary orbits, we tested the habitability of each planet assuming an orbit around each stellar component separately, as we currently lack data indicating which stars host which (or any) planets in these systems.

The existence of other bright stars in the \kepler photometric aperture (in this case due to the stellar multiplicity of the systems) required that the recorded transit depth be corrected for the light dilution from the additional star(s). To account for the transit dilution, we scaled the blended transit depth observed by \kepler by the photometric contribution of the star of interest, as

\begin{equation}
	\label{eq:dilution}
	\Delta F_{true} ={\Delta F_{\rm MAST}}/{dilution}
\end{equation}

\noindent where ${\Delta F_{\rm MAST}}$ is the transit depth as measured by {\it Kepler}, and {\it dilution} is the fraction of the blended light in the \kepler aperture that is contributed by the individual stellar components. The dilutions to the transit depth were calculated using the PSF fitting (\S\ref{sec:psf}) coupled with the $\rm{Kp-{\it HST}}$ conversion (\S\ref{sec:kphst}), and are listed in \S\ref{sec:psf}. As each star is smaller and cooler than the raw \kepler photometry indicates (as \kepler only shows the blended system), the relative drop in the stellar flux due to the transit is actually larger than was measured, which in turn increases the ratio of ${\rm R_p/R_*}$. The input transit parameters used in the habitability calculations are found in Table \ref{tab:transit}. The errors listed for ${\Delta F_{true}}$ were calculated using the detection S/N and the archive-listed transit depth in parts per million. 

\begin{table}[t]
	\begin{center}
	\caption{Transit Parameters for {\it Kepler}-296, \triple, and \newbin\,Components}
	\label{tab:transit} 
		\begin{tabular*}{0.5\textwidth} {@{\extracolsep{\fill} } l c c c} 
			\hline \hline
			Planet\tablenotemark{a} 	&	{$\Delta \rm{F_{\rm MAST}}$} \tablenotemark{b}	& 	${\Delta \rm{F_{true}}}$ \tablenotemark{c}  	   	&	Period \tablenotemark{b} \\  
								& 	[ppm]					& 	[ppm]								& [days]\\ \hline
			\binary\,Ac			&	$1423.0\pm28.1$			&$1767.7\pm	34.9	$					&	5.842 \\ 
			\binary\,Ad			&	$1567.0\pm41.2$			&$1946.6\pm	51.2	$					&	19.850\\ 
			\binary\,Ab			&	$820.0\pm36.3$			&$1018.6\pm	45.1	$					&	10.864\\ 
			\binary\,Af				&	$979.0\pm60.8$			&$1216.1\pm	75.5	$					&	63.338\\ 
			\binary\,Ae			&	$787.0\pm45.8$			&$977.6\pm	56.8	$					&	34.142\\ \hline
			\binary\,Bc			&	$1423.0\pm28.1$			&$7297.4\pm	143.9$					&	5.842\\ 
			\binary\,Bd			&	$1567.0\pm41.2$			&$8035.9\pm	211.5$					&	19.850\\ 
			\binary\,Bb			&	$820.0\pm36.3$			&$4205.1\pm	186.1$					&	10.864\\ 
			\binary\,Bf				&	$979.0\pm60.8$			&$5020.5\pm	311.8$					&	63.338\\
			\binary\,Be			&	$787.0\pm45.8$			&$4035.9\pm	234.6$					&	34.142\\ \hline
			\triple\,A.01			&	$818.0\pm47.3$			&$1506.4\pm	87.1	$					&	38.098\\ 	
			\triple\,B.01			&	$818.0\pm47.3$			&$2690.8\pm	155.5$					&	38.098\\ 
			\triple\,C.01			&	$818.0\pm47.3$			&$5346.4\pm	309.0$					&	38.098\\ \hline
			\newbin\,A.01			&	$540.0\pm32.0$			&$866.8\pm	51.3	$					&	22.477\\ 
			\newbin\,B.01			&	$540.0\pm32.0$			&$1432.4\pm	84.8	$					&	22.477\\ \hline
		\end{tabular*}
		\end{center}
		\vspace{-0.2cm}
	\tablenotetext{1}{``\binary\,A{\it c}" etc. indicates the solution for planet c around component A of \binary.}
	\tablenotetext{2}{From MAST.} 
	\tablenotetext{3}{Corrected for dilution from the stellar companion via Eq. \ref{eq:dilution}.} 

\end{table}

\subsection{Calculation of Planetary Parameters}
\label{sec:eqns}

Using the transit parameters listed in Table \ref{tab:transit}, we calculated the planet radius, the semi-major axis, the equilibrium temperature, and incident stellar flux of each planet around each of its potential host stars using the equations listed in \cite{seager03}. Planetary masses and bulk densities were calculated using the formalisms of \cite{weiss14} and \cite{liss11}. These formalisms do not take into account stellar limb darkening, instead assuming a uniform stellar disk. {We provide these results as a first order calculation, and provide the results of limb darkened model fits to the full folded time series in the next subsection.} 

The planetary radius was directly calculated from the stellar radius and the transit depth using the equations of \cite{seager03}, as
\begin{equation}
	R_p=R_\star \sqrt{\Delta F_{true}}
	\label{eq:radius}
\end{equation}

\noindent where ${\Delta F_{true}}\,$ is the dilution-corrected transit depth from Eq. \ref{eq:dilution} and ${R_\star }$ is the stellar radius. The {planetary} orbital semi-major axis was calculated from the KIC transit period and the best-fit stellar mass, using

\begin{equation}
	a_p=a_\oplus \left(\frac{P_p}{P_\oplus} \right)^{2/3} \left( \frac{M_\star}{M_\odot}\right ) ^{1/3}
	\label{eq:semimajor}
\end{equation}

\noindent where ${P_p}$ is the planetary orbital period and ${M_\star}$ is the stellar mass. The semi-major axis calculated in Eq. \ref{eq:semimajor} was combined with the best-fit stellar effective temperature and radius to get the planetary equilibrium temperature via

\begin{equation}
	T_{eq}= \teff {(1-A )^{1/4}  \sqrt{\frac{R_\star}{2\,a_p}   }	}	
	\label{eq:teq}
\end{equation}

\noindent where $A$ is the assumed Bond albedo of 0.3 and $a_p$ is the planetary semi-major axis as calculated in Eq. \ref{eq:semimajor}. This equilibrium temperature does not account for any potential greenhouse effects, which {would warm the surface and are unavoidable} if there is any liquid water on the surface. Next, the stellar flux incident on the planet was calculated relative to the flux received at Earth by

\begin{equation}
	\frac{S_{\rm eff}}{S_0} = \left(\frac{1\au}{a_p}\right)^2 \left(\frac{R_\star}{\Rsun}\right)^2 \left(\frac{T_*}{\rm T_\odot}\right)^4  
	\label{eq:fpfe}
\end{equation}

\noindent where ${a_p}$ is the planetary semi-major axis, ${R_\star }$ is the stellar radius, $T_*$ is the stellar temperature, and $T_\odot = 5779{\,\rm K}$ is the adopted value of solar effective temperature.

Lastly, the mass and density of the planets were calculated using the empirical relations of \cite{weiss14} for planets less than 4 Earth-radii, given as

\begin{equation}
	\rho_p = 2.43+3.39\left(\frac{R_p}{R_\oplus}\right)\,{\rm g/cm^3}
	\label{eq:density}
\end{equation}

\noindent for $R_p/R_\oplus < 1.5$ and 

\begin{equation}
	\frac{M_p}{M_\oplus} = 2.69\left(\frac{R_p}{R_\oplus}\right)^{0.93}\,{\rm g/cm^3} 
	\label{eq:mass}
\end{equation}

\noindent for $1.5\leq R_p/R_\oplus < 4$. The relation of \cite{liss11} was used for planets  with $R_p/R_\oplus \geq4$, as

\begin{equation}
	M_{p} = \left(\frac{R_p}{R_\oplus}\right)^{2.06}\,M_\oplus
	\label{eq:lissmass}
\end{equation}

\noindent which fits exoplanet observations for planets smaller than Saturn. Conversion between mass and density was done using

\begin{equation}
	\frac{\rho_p}{\rho_\oplus} = \frac{M_p/M_\oplus}{\left(R_p/R_\oplus	\right)^3}
	\label{eq:massdensity}
\end{equation}

We used the formalism of \cite{ravihz} to determine the habitability of the planets. Using Eq. 2 from that paper, we calculated the locations of the moist greenhouse limit (inner) and the maximum greenhouse limit (outer) for each of our component stars and compared the limits to the calculated effective stellar flux incident on the planets from Eq. \ref{eq:fpfe}. If a planet falls between the moist and maximum greenhouse limits, we considered it to be habitable. The moist and maximum greenhouse limits were chosen to be conservative locations of the Habitable Zone, though for stars with $\teff\lesssim5000{\rm\, K}$ the moist greenhouse limit is indistinguishable from the runaway greenhouse limit.

The projected separations of the stellar components in both systems range from $\sim50-225\au$, while the orbital periods of the planets as measured by \kepler are on the order of weeks. The wide separations of the components of each system greatly reduce the chances that the {stellar} components produce overlapping Habitable Zones like in close (i.e. $<50\au$) multi-star systems \citep{stype13}. Furthermore, censuses of the populations of protoplanetary disks in wide ($\gtrsim 40\au$) binary systems show that the influence of a binary companion reduces the lifetime of the disk by a few Myr, which decreases the likelihood of planet formation \citep{kraus12}. As these systems successfully completed planet formation, the protoplanetary disk was likely only affected minimally by the stellar companion(s), {further }suggesting independent Habitable Zones. 

\subsection{Transit Light Curve Fitting} 
\label{sec:transits}
 
{
The above evaluation of planet habitability in each system is accurate to first order, but the equations in \S\ref{sec:eqns} do not account for stellar limb darkening, orbital eccentricity, inclination, or impact parameter. These exclusions affect our calculation of the planetary radius and mass, and thus could potentially change our conclusions about planetary habitability. We adopted a more robust method of transit analysis by fitting a transit model using an MCMC algorithm to iteratively solve for the best fitting transit model.  Attempts at using publicly available MCMC transit fitting software, including the Transit Analysis Package {\citep[TAP;][]{tap12}},  EXOFAST \citep{eastman}, and PyKE packages \citep{pyke}, illuminated limitations in dealing with low mass and low stellar temperature cases. We found that the transit identifying function {\tt autokep} built in to TAP was unable to identify the transits of these systems without first stitching together light curves from all of the quarters, folding them on their linear ephemerides, and binning the phase-folded light curve using PyKE packages. The EXOFAST transit fitter, attempted first through the TAP GUI and then use of the function directly, showed that their stellar mass-radius relation \citep{torres} was unable to handle stellar masses below $0.6\,\Msun$ and that their limb-darkening interpolation functions were unsupported for stellar temperatures below 3500 K. While tests using EXOFAST showed that the transit solutions for $M_\star>0.6\Msun$, $\teff>3500{\textrm K}$ transits were reliable, the mass and temperature limits imposed by the program during execution were unsuitable for the stellar solutions in this study.

We modified both the EXOFAST code itself and the input transit light curves. We applied an adaptive binning algorithm to the input transit light curves to ensure that the transit itself was properly sampled. This properly preserved the shape and depth of the transits while reducing computation time with broader bins outside of transit. We took the mean time of all the data points within a bin as the bin time value, rather than the bin midpoint, to account for any clumps or gradients within a bin and aid in accurate reproduction of transit shape. We used Poisson statistics to calculate the uncertainty in the mean flux value of each bin; this led to smaller uncertainties in the out-of-transit points and larger uncertainties within the transit, which allowed EXOFAST to properly weight each binned flux value. Finally, after binning the light curves for each planet in our sample, we applied the stellar dilution corrections directly to the light curves themselves using Eq. \ref{eq:dilution} as before. This produced a separate light curve for each possible planet/star permutation. EXOFAST was then used in a mode that integrates the \cite{mandel} light curve model over a long cadence period (29.4 minutes), a smoothing to the data that applies even when binning within transits to shorter intervals.

Within the EXOFAST package itself, we overrode the built-in stellar mass-radius relation from \cite{torres} since the function was unreliable when extrapolated to stellar masses below $0.6\,\Msun$. As we wanted to enforce our isochrone solutions for the stellar mass and radius, we imposed those solutions as prior values and calculated the prior widths from our uncertainties in the stellar mass and radius solutions. We then added a penalty to the \chisq calculation within EXOFAST for deviating from the desired stellar mass and radius. The uncertainties in the stellar mass and radius from the isochrone fitting are then accuratly propagated through EXOFAST into the posterior distributions and resulting uncertainties for the planetary values. We utilized the online limb darkening applet from \cite{eastman} to calculate stellar limb darkening priors for our transit fitting to support calculation of limb darkening coefficients for stellar temperatures below 3500K. The online limb darkening utility interpolates the quadratic limb darkening tables of \cite{limbdark} given a bandpass, effective temperature, surface gravity, and stellar metallically. We calculated the quadratic limb darkening separately and imposed those values as additional priors with small prior widths. In addition to priors on the stellar properties, the planetary orbital period, and transit center time, we included a prior restriction on the orbital eccentricity to downweight high eccentricity solutions that are unphysical and skew the posterior distributions of all related variables.

We applied these modifications to EXOFAST and the input transit light curves and then fit transit models to the light curves for each possible permutation of planet and star as done previously with the analytic solutions. Before accepting the EXOFAST solution as ``good", we assured that the reduced \chisq\, of the transit fit was $\sim1$, that the best fit stellar parameters indicated by EXOFAST (especially the stellar effective temperature) matched our isochrone solutions within $1\sigma$, and that the calculated $R_P/R_*$ matched the value calculated analytically in Eq. \ref{eq:radius}. As the MCMC fitting did not account for the observed {\it HST} photometry which constrained our stellar solutions, these checks ensured that the MCMC algorithm did not diverge from the isochrone fits or indicate a solution that was not consistent with observations. 


}

\subsection{Implications on Habitability} 
\label{sec:hab}

\begin{table*}[t!]
	\begin{center}
	\caption{Analytic {and EXOFAST Solutions} for {\it Kepler}-296, \triple, and \newbin\, Planets}
	\label{tab:planets} 
		\begin{tabular*}{\textwidth} {@{\extracolsep{\fill} } c c c c c c c c} 
			\hline \hline
Planet\tablenotemark{a} 	&	${R_p}$ 			&	{${a_P}$} 	&	${\rm M_p}$	&	$\rho_p$		&	{$\teq$} 		&	{${S_{\rm eff}}$}	&	{HZ}\tablenotemark{b}		\\ 
					&	${[R_\oplus]}$	&	[AU]		&	${[M_\oplus]}$	&	${\rm [g/cm^3]}$	&	[K]			&	${[S_0]}$			&		\\ \hline
\binary\, A{\it c}			&	$2.75\pm0.33$	&	0.054	&	6.9			&	1.8			&	$558.6\pm41.0$	&	$22.92\pm6.73$	&	no\\
					&	$3.35\pm0.21$	&	0.054	&	8.3			&	1.2			&	$606.0\pm32.0$	&	$22.63\pm2.20$	&	\emph{no}\\
\binary\, A{\it d} 			&	$2.88\pm0.35$	&	0.123	&	7.2			&	1.7			&	$371.5\pm27.3$	&	$4.49\pm1.32$		&	no\\
					&	$2.69\pm0.21$	&	0.123	&	6.8			&	1.9			&	$403.0\pm21.5$	&	$4.26\pm0.98$		&	\emph{no}\\
\binary\, A{\it b}			&	$2.09\pm0.26$	&	0.082	&	5.3			&	3.2			&	$454.2\pm33.3$	&	$10.02\pm2.94$	&	no\\
					&	$2.15\pm0.21$	&	0.082	&	5.5			&	3.0			&	$495.0\pm25.5$	&	$10.07\pm4.58$	&	\emph{no}\\
\binary\, A{\it f}			&	$2.28\pm0.28$	&	0.266	&	5.8			&	2.7			&	$252.4\pm18.5$	&	$0.95\pm0.28$		&	maybe\\
					&	$2.08\pm0.21$	&	0.266	&	5.3			&	3.2			&	$274.0\pm15.0$	&	$0.88\pm0.46$		&	\emph{\bf yes}\\
\binary\, A{\it e}			&	$2.04\pm0.25$	&	0.176	&	5.2			&	3.4			&	$310.1\pm22.8$	&	$2.18\pm0.64$		&	no\\		
					&	$1.86\pm0.17$	&	0.176	&	4.8			&	4.1			&	$337.0\pm17.5$	&	$2.04\pm0.62$		&	\emph{no}\\	\hline
\binary\, B{\it c}			&	$4.03\pm0.68$	&	0.049	&	17.7			&	1.5			&	$450.3\pm42.9$	&	$9.68\pm3.69$		&	no\\
					&	$3.78\pm0.45$	&	0.049	&	9.3			&	0.9			&	$497.0\pm27.0$	&	$9.99\pm1.48$		&	\emph{no}\\
\binary\, B{\it d}			&	$4.23\pm0.71$	&	0.110	&	19.5			&	1.4			&	$299.5\pm28.6$	&	$1.89\pm0.72$		&	no\\
					&	$4.00\pm0.45$	&	0.110	&	17.4			&	1.5			&	$331.0\pm21.5$	&	$1.98\pm0.71$		&	\emph{no}\\
\binary\, B{\it b}			&	$3.06\pm0.52$	&	0.074	&	7.6			&	1.5			&	$366.1\pm34.9$	&	$4.23\pm1.61$		&	no\\
					&	$2.91\pm0.63$	&	0.074	&	7.3			&	1.6			&	$395.0\pm33.0$	&	$3.82\pm1.12$		&	\emph{no}\\
\binary\, B{\it f}			&	$3.35\pm0.57$	&	0.239	&	8.3			&	1.2			&	$203.4\pm19.4$	&	$0.40\pm0.15$		&	{\bf yes}\\
					&	$2.78\pm0.40$	&	0.240	&	7.0			&	1.8			&	$214.0\pm16.5$	&	$0.34\pm0.31$		&	\emph{\textbf{yes}}\\
\binary\, B{\it e}	 		&	$3.00\pm0.51$	&	0.158	&	7.5			&	1.5			&	$250.0\pm23.7$	&	$0.92\pm0.35$		&	maybe\\	
					&	$2.72\pm0.38$	&	0.158	&	6.8			&	1.9			&	$273.0\pm17.5$	&	$0.91\pm0.48$		&	\emph{maybe}\\\hline
\triple\, A.01 			&	$2.04\pm0.33$	&	0.176	&	5.2			&	3.4			&	$265.6\pm24.2$	&	$1.17\pm0.43$		&	maybe\\
					&	$1.86\pm0.25$	&	0.176	&	4.8			&	4.1			&	$289.0\pm20.0$	&	$1.13\pm0.58$		&	\emph{maybe} \\
\triple\, B.01 			&	$2.37\pm0.44$	&	0.168	&	6.0			&	2.5			&	$244.6\pm25.2$	&	$0.84\pm0.35$		&	{\bf yes}\\
					&	$2.47\pm0.35$	&	0.176	&	6.2			&	2.3			&	$278.0\pm18.5$	&	$0.99\pm0.53$		&	\emph{maybe}\\
\triple\, C.01 			&	$2.58\pm0.62$	&	0.153	&	6.5			&	2.1			&	$216.9\pm27.6$	&	$0.52\pm0.27$		&	{\bf yes}\\
					&	$2.65\pm0.28$	&	0.150	&	6.6			&	2.0			&	$252.0\pm13.0$	&	$0.68\pm0.37$		&	\emph{\textbf{yes}}\\	\hline
\newbin\, A.01 			&	$1.90\pm0.24$	&	0.132	&	4.9			&	3.9			&	$422.1\pm29.8$	&	$7.47\pm2.11$		&	no\\
					&	$1.57\pm0.10$	&	0.132	&	4.1			&	5.8			&	$461.0\pm20.5$	&	$7.57\pm1.17$		&	\emph{no}\\
\newbin\, B.01 			&	$2.23\pm0.30$	&	0.128	&	5.7			&	2.8			&	$386.1\pm29.4$	&	$5.23\pm1.60$		&	no \\
					&	$1.97\pm0.17$	&	0.128	&	5.1			&	 3.6			&	$436.0\pm22.0$	&	$5.88\pm1.10$		&	\emph{no}\\
			
  \hline
		\end{tabular*}
		\end{center}
		\vspace{-0.15cm}
		\tablecomments{The first row for each planet contains the analytic planet solution and the second row for each planet contains the EXOFAST planet solution. The HZ determination is italicized for the EXOFAST solution and bolded for any HZ planets.}
		\tablenotetext{1}{The notation ``\binary\,A{\it c}" etc. indicates the solution for planet c around component A of \binary.}
		\tablenotetext{2}{HZ indicates falling between the moist greenhouse inner limit and max greenhouse outer limit. {``maybe" indicates falling within 1$\sigma$ of the HZ.}}
\end{table*}

Table \ref{tab:planets} lists the calculated planetary parameters for each planet around each potential stellar host {for both the analytic method and the EXOFAST method. The tabulated EXOFAST solutions are the median values and the 68\% confidence intervals on the posterior MCMC distributions}. We find planetary radii that range from  ${1.57}\Rearth$ to $4.23\Rearth$ and are larger than those listed in the Mikulski Archive for Space Telescopes\footnote{\url{http://archive.stsci.edu/}} (MAST) due to the dilution corrections. Regardless of the host star around which the planets orbit, all planets around \binary\, and the single planets around \triple\,and \newbin\, are super-Earths/mini-Neptunes. Our calculated values of planetary radius are larger than those tabulated in \cite{dres13} and \cite{muir12} for \binary\,{\it c}, \binary\,{\it d}, and \binary\,{\it b}, and larger than the radii recorded in MAST for all planets in the \binary\,system due to our inclusion of the transit depth dilution. Our planetary radius for \triple.01 is also larger than those recorded in MAST and \cite{dres13}, and our radius for \newbin.01 is larger than the MAST value for the same reason.

{Upon comparison of the analytic and EXOFAST solutions, we note that the planetary radius (rather, $R_p/R_*$ in the calculation) and the effective stellar flux are mildly dependent on the inclusion of limb darkening, and consequently the planetary mass and equilibrium temperatures are also mildly dependent on the inclusion of higher order calculations. As expected, planets that fall in the HZ according to the analytic solutions are still habitable with the EXOFAST calculations, either falling directly within the HZ or within $1\sigma$ of the inner edge of the HZ.}

Figure \ref{fig:hzplanets} displays a subset of planets that fall in or near the Habitable Zones of their potential host star {according to the EXOFAST solutions} and helps highlight the differences between our calculations and those of of \cite{dres13} and \cite{muir12}. Both \citeauthor{dres13} and \citeauthor{muir12} determined that \binary\,{\it d} was in the Habitable Zone of the assumed single star. Using our stellar solutions for \binary, \binary\,{\it d} is not habitable around either star, and in fact falls significantly interior to the Habitable Zone of either star. {The outermost planet in the system (\binary\,{\it f}) now falls comfortably within the Habitable Zones of both the primary and the secondary stars. \binary\,{\it e} also falls just barely interior to the Habitable Zone of the secondary, but the uncertainty on the effective stellar flux at that planet makes it another likely habitable candidate.} Neither \citeauthor{dres13} nor \citeauthor{muir12} reported on the status of \binary\,{\it f} or \binary\,{\it e} due to the timing of the two studies.

{The multiplicity of \triple\, also changes our understanding of the habitability of its single planet. \citeauthor{dres13} report that \triple.01 falls within the Habitable Zone of the assumed single star, but our results show that this is only possible around the tertiary star. The uncertainty in the effective stellar flux indicates that \triple.01 may also be habitable around the primary and secondary stars despite its location interior to the HZ. }

Lastly, we find that the multiplicity of \newbin\, does not improve its planet's chances of habitability. Even with the stellar dilution to the transit depth accounted for, \newbin.01 remains well interior to the Habitable Zone around both the primary and secondary components, as it also did for the initial single-star analysis.


\begin{figure}[t]
	\begin{center}
	\includegraphics[width=0.45\textwidth]{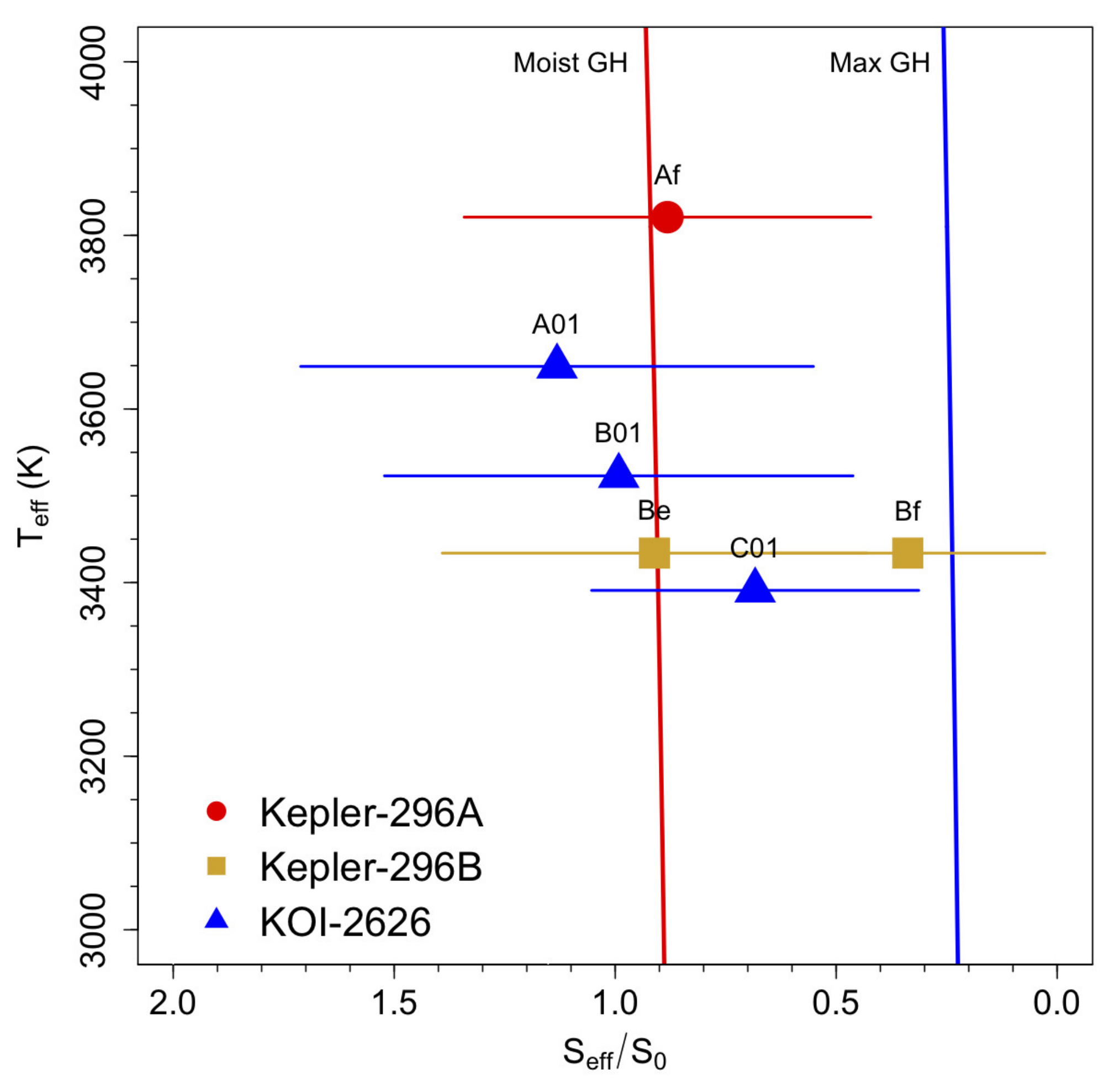}
	\caption{Stellar effective temperature versus effective {incident stellar} flux {from EXOFAST} in solar units for planets in and near the Habitable Zones of their respective stars. Red circles indicate \binary\,A, gold squares indicate \binary\,B, and blue triangles indicate \triple. Moist and max greenhouse curves are calculated using formalism of \cite{ravihz}. Any planets not shown fall significantly interior to the Habitable Zone. Planet labels as in Table \ref{tab:transit}.}    
	\label{fig:hzplanets}
	\end{center}
\end{figure}
\section{Discussions and Future Work}
\label{sec:discussion}

\cite{dres13} report a temperature for the blended \binary\, of $3424\pm50\rm{\, K}$, while \cite{muir12} report a temperature of $3517 \rm{\,K}$ based on spectral index matching. Our best-fit isochrone temperatures for both components A and B are warmer than the \citeauthor{dres13} values. However, our temperatures do straddle the blended temperature of \cite{muir12} as expected. \cite{mann13} report $\teff=3622\rm{\,K}$ for \binary, which also falls between our temperatures of the individual components as expected. Likewise for \triple, \cite{dres13} adopt a value of $\teff = 3482\rm{\,K}$, which falls between our values for components B and C, while \cite{mann13} report $\teff=3637\rm{\,K}$ which falls between our solutions for components A and B. That our solutions agree with blended temperature estimates derived using two different methods suggests that the VR isochrones provided a logical solution for both \binary\, and \triple. \cite{muir12} did not include the \triple\, system in their studies, and none of the aforementioned reports included \newbin.

Our initial analysis attempted to follow the procedure outlined in earlier sections of this paper, but utilizing the {DSED} isochrones in place of the VR isochrones. This was initially an attempt to best compare to the studies of \cite{dres13} and \cite{muir12}, the former of which also fit to Dartmouth isochrones and the latter which produced consistent results using spectroscopic methods. Our first results from using the Dartmouth isochrones indicated temperatures for all components that were much hotter than the temperatures reported by both studies (and later reported by \cite{mann13} as well). Investigating the cause of this difference, we attempted first to replicate the results of \cite{dres13} regarding the temperature of \binary, using the same seven bands that were used in that study ($grizJHK$). We were able to match the \cite{dres13} $\teff$ to within 100 K, and found that the inclusion on the SDSS {\it g} band photometry skewed the isochrone fitting to significantly cooler temperatures. Dropping the {\it g} band photometry produced a warmer midpoint between A and B temperatures and a large drop of \chisq, while exclusion of any other band made little difference on the temperature midpoint or \chisq. Knowing {\it a priori} the late spectral types of the targets, we observe that the inclusion of {\it g} band photometry may bias some of the isochrone solutions of \citeauthor{dres13}. Photometry in the {\it g} band is also observationally suspect in the KIC at those faint magnitudes \citep{kic11}. The photometric issues are then coupled with the uncertainties of the Dartmouth isochrones for late-type stars as discussed in \S\ref{sec:isochrone}. We also note that our analysis is limited to the use of optical and near-optical bandpasses, which are not the most reliable wavelength ranges for cooler stars. {To mitigate this we relied more heavily on our NIR bandpass over our optical bandpass when fitting our photometry to the VR isochrones.} Inclusion of infrared bands for these targets will likely affect the temperatures derived from the isochrone fitting and {reduce} the differences between VR and Dartmouth isochrones.



Habitable planets in the canonical sense must not only have the capability for liquid water on the surface, but also have a solid surface on which that water can exist. In short, the planets must be rocky and not gaseous. {Using radial velocity measurements coupled with Doppler spectroscopy, high-resolution imaging, and asteroseismology, \citet{marcy13} measured the radii and masses for 65 planet candidates and concluded that only planets with radii less than $\sim {1.5}\,R_\oplus$ are compatible with purely rocky compositions. Planets larger than that must have a larger fraction of low-density material, e.g. H, He, and $\rm{H_2 O}$. Our updated planet radii from EXOFAST indicate that none of our potentially habitable planets (\binary\,A{\it f}, \binary\,B{\it f}, \binary\,B{\it e}, \triple\,A.01, \triple\,B.01, and \triple\,C.01) are small enough to have purely rocky compositions according to \cite{marcy13}, and thus are not habitable in the canonical sense. \newbin\,A.01, however, is within $1\sigma$ of the purely rocky composition limit and so may still be a rocky planet.} We cannot exclude the possibility of a very massive yet rocky planet like {\it Kepler}-10c \citep{megaearth} as we lack radial velocity measurements needed to calculate the planetary masses and densities directly. {Even if \binary\,A{\it f}, \binary\,B{\it f}, \binary\,B{\it e}, \triple\,A.01, \triple\,B.01, and \triple\,C.01 remain} too large to be rocky, the possibility of habitable exomoons would remain.

\section{Conclusion}
\label{conclusion}
Using the results of our {\it HST} GO/SNAP program \pid\, we derived {\it HST}-based photometry for the hosts of some of the most interesting \kepler planet candidates and created a conversion between the broad-band Kp and our two filters from {\it HST}. We utilized the empirical PSF from \cite{gs14} for \binary, \triple, and \newbin, three \kepler targets that were recently discovered to be tight multi-star systems with small and cool planets. Based on the goodness of the binary isochrone fitting, we determined that components A and B in \binary\, are almost certainly a bound, coeval system consisting of two early-M dwarfs. Based on the updated stellar properties from the Victoria-Regina Stellar Model isochrone matches, we found that the system still contains {a potentially habitable planet around its primary star and two potentially habitable planets around its secondary star}, with all other combinations of star-planet producing too-hot planets. Likewise, we found that \triple\, is likely a bound, coeval, triple star system containing three early- to mid-M dwarfs with a single planet that is potentially habitable around {any of the stellar components.} Lastly, while \newbin\, is likely also a bound, binary K dwarf system, its single planet is not habitable around either stellar component. {While the sizes of \binary\,A{\it f}, \binary\,B{\it f}, \binary\,B{\it e}, \triple\,A.01, \triple\,B.01, and \triple\,C.01 indicate that those planets are most likely gaseous, \newbin\,A.01 likely has a mostly rocky compositions based on the work of \cite{marcy13}, though it is well interior to the HZ of its star. The six potentially habitable planets have densities more consistent with a higher gaseous fraction and are not likely habitable in the canonical sense.} 

\acknowledgements
K.M.S.C. performed analyses found in \S\ref{sec:obs}, \S\ref{sec:multistar}, and \S\ref{sec:planets} and discussion in \S\ref{sec:intro}, \S\ref{sec:discussion}, and \S\ref{conclusion}. R.L.G. contributed analysis to \S\ref{sec:psf} and \S\ref{sec:fp} as well as overall guidance and direction for this work and the companion paper \cite{gs14} J.T.W. contributed to \S\ref{sec:intro}, \S\ref{conclusion}, and valuable discussion and advice regarding isochrone use. D.R.C. contributed Keck AO K-band data to \S\ref{sec:2626fit} and provided discussion on \triple. K.M.S.C. and R.L.G. have been partially supported through grant {\it HST}-GO-12893.01-A from STScI. We thank Don VandenBerg for permitting use of the latest Victoria-Regina Stellar Models before publication. We also thank Sharon X. Wang for discussion on error analysis for our isochrone fitting.

Some of the data presented in this paper were obtained from the Mikulski Archive for Space Telescopes (MAST). STScI is operated by the Association of Universities for Research in Astronomy, Inc., under NASA contract NAS5-26555. Support for MAST for non-{\it HST} data is provided by the NASA Office of Space Science via grant NNX13AC07G and by other grants and contracts. This paper makes use of data collected by the \kepler mission. Funding for the \kepler mission is provided by the NASA Science Mission directorate. Some of the data presented herein were obtained at the W.M. Keck Observatory, which is operated as a scientific partnership among the California Institute of Technology, the University of California and the National Aeronautics and Space Administration. The Observatory was made possible by the generous financial support of the W.M. Keck Foundation. The Center for Exoplanets and Habitable Worlds is supported by the Pennsylvania State University, the Eberly College of Science, and the Pennsylvania Space Grant Consortium.We gratefully acknowledge the use of SOA/NASA ADS, NASA, and STScI resources. \\ \\


Facilities: \facility{{\it HST} (WFC3)}, \facility{{\it Kepler}}.
\\ 



\begin{thebibliography}{30}
\expandafter\ifx\csname natexlab\endcsname\relax\def\natexlab#1{#1}\fi

{\bibitem[Allard et al.(2011)]{btsettl} Allard, F., Homeier, D., \& Freytag, B.\ 2011, 16th Cambridge Workshop on Cool Stars, Stellar Systems, and the Sun, 448, 91}
%
\bibitem[Batalha et al.(2013)]{batalha13} Batalha, N.~M., Rowe, J.~F., Bryson, S.~T., et al.\ 2013, \apjs, 204, 24 
%
\bibitem[Borucki et al.(2010)]{borucki10} Borucki, W.~J., Koch, D., Basri, G., et al.\ 2010, Science, 327, 977
%
\bibitem[Borucki et al.(2011)]{borucki11} Borucki, W.~J., Koch, D.~G., Basri, G., et al.\ 2011, \apj, 736, 19 
%
\bibitem[Boyajian et al.(2012)]{kdwarf} Boyajian, T.~S., von Braun, K., van Belle, G., et al.\ 2012, \apj, 757, 112 
%
\bibitem[Brown et al.(2011)]{kic11} Brown, T.~M., Latham, D.~W., Everett, M.~E., \& Esquerdo, G.~A.\ 2011, \aj, 142, 112 
%
\bibitem[Burke et al.(2014)]{burke14} Burke, C.~J., Bryson, S.~T., Mullally, F., et al.\ 2014, \apjs, 210, 19
%
%
\bibitem[Cantrell et al.(2013)]{recons13} Cantrell, J.~R., Henry, T.~J., \& White, R.~J.\ 2013, \aj, 146, 99
%
{\bibitem[Casagrande \& VandenBerg(2014)]{vdb2} Casagrande, L., \& VandenBerg, D.~A.\ 2014, \mnras, 444, 392}
%
%
\bibitem[Claret \& Bloemen(2011)]{limbdark} Claret, A., \& Bloemen, S.\ 2011, \aap, 529, AA75
%
\bibitem[Croll et al.(2014)]{croll13} Croll, B., Rappaport, S., 
DeVore, J., et al.\ 2014, \apj, 786, 100
%
\bibitem[Dotter et al.(2008)]{dsed08} Dotter, A., Chaboyer, B., Jevremovi{\'c}, D., et al.\ 2008, \apjs, 178, 89
%
\bibitem[Dressing 
\& Charbonneau(2013)]{dres13} Dressing, C.~D., \& Charbonneau, D.\ 2013, \apj, 767, 95 
%
%
\bibitem[Dumusque et al.(2014)]{megaearth} Dumusque, X., Bonomo, 
A.~S., Haywood, R.~D., et al.\ 2014, \apj, 789, 154
%
\bibitem[Eastman et al.(2013)]{eastman} Eastman, J., Gaudi, B.~S., \& Agol, E.\ 2013, \pasp, 125, 83
%
\bibitem[Feiden et al.(2011)]{feiden11} Feiden, G.~A., Chaboyer, 
B., \& Dotter, A.\ 2011, \apjl, 740, L25 
%
%
\bibitem[Fressin et al.(2013)]{fressin13} Fressin, F., Torres, 
G., Charbonneau, D., et al.\ 2013, \apj, 766, 81
%
{\bibitem[Frutcher et al.(2010)]{adriz} Fruchter, A.S., Hack, W., Dencheva, N., Droettboom, M., Greenfield, P. 2010, STSCI Calibration Workshop Proceedings, Baltimore, MD, STScI, 376}
%
\bibitem[Fukugita et al.(1996)]{sdss96} Fukugita, M., 
Ichikawa, T., Gunn, J.~E., et al.\ 1996, \aj, 111, 1748
%
\bibitem[Gazak et al.(2012)]{tap12} Gazak, J.~Z., Johnson, J.~A., Tonry, J., et al.\ 2012, Advances in Astronomy, 2012 
%
\bibitem[Gilliland 
\& Rajan(2011)]{gill11} Gilliland, R.~L., \& Rajan, A.\ 2011, Instrument Science Report WFC3 2011-03 (Baltimore, MD: STScI) 
%
\bibitem[Gilliland et al.(2015)]{gs14} Gilliland, R.~L., 
Cartier, K.~M.~S., Adams, E.~R., et al.\ 2015, \aj, 149, 24 
%
\bibitem[Gonzaga et al.(2012)]{drizzlepac}Gonzaga, S., Hack, W., Fruchter, A., \& Mack, J. 2012, The DrizzlePac Handbook, Baltimore, STScI
 %
 \bibitem[Hauschildt et al.(1999a)]{phoenixa} Hauschildt, P.~H., Allard, F., \& Baron, E.\ 1999, \apj, 512, 377
 %
 \bibitem[Hauschildt et al.(1999b)]{phoenixb} Hauschildt, P.~H., Allard, F., Ferguson, J., Baron, E., \& Alexander, D.~R.\ 1999, \apj, 525, 871 
 %
 \bibitem[Henry et al.(1999)]{recons99} Henry, T.~J., Franz, O.~G., Wasserman, L.~H., et al.\ 1999, \apj, 512, 864
%
\bibitem[Henry et al.(2006)]{recons06} Henry, T.~J., Jao, W.-C., Subasavage, J.~P., et al.\ 2006, \aj, 132, 2360
%
{\bibitem[Hinkel et al.(2014)]{hypatia} Hinkel, N.~R., Timmes, F.~X., Young, P.~A., Pagano, M.~D., \& Turnbull, M.~C.\ 2014, \aj, 148, 54 }
%
\bibitem[Howard et al.(2012)]{howard12} Howard, A.~W., Marcy, G.~W., Bryson, S.~T., et al.\ 2012, \apjs, 201, 15 
%
\bibitem[Jao et al.(2014)]{recons14} Jao, W.-C., Henry, T.~J., Subasavage, J.~P., et al.\ 2014, \aj, 147, 21
%
\bibitem[Kaib et al.(2013)]{kaib13} Kaib, N.~A., Raymond, S.~N., \& Duncan, M.\ 2013, \nat, 493, 381 
%
\bibitem[Kaltenegger \& Haghighipour(2013)]{stype13} Kaltenegger, L., \& Haghighipour, N.\ 2013, \apj, 777, 165 
%
\bibitem[Kasting et al.(1993)]{kasthz} Kasting, J.~F., Whitmire, D.~P., \& Reynolds, R.~T.\ 1993, Icarus, 101, 108 
%
\bibitem[Kopparapu(2013)]{dresrevise} Kopparapu, R.~K.\ 2013, \apjl, 767, L8 
%
\bibitem[Kopparapu et al.(2013)]{ravihz} Kopparapu, R.~K., Ramirez, R., Kasting, J.~F., et al.\ 2013, \apj, 765, 131 
%
\bibitem[Kratter \& Perets(2012)]{kratter12} Kratter, K.~M., \& Perets, H.~B.\ 2012, \apj, 753, 91 
%
\bibitem[Kraus et al.(2012)]{kraus12} Kraus, A.~L., Ireland, M.~J., Hillenbrand, L.~A., \& Martinache, F.\ 2012, \apj, 745, 19 
%
\bibitem[L{\'e}pine et al.(2013)]{lepine13} L{\'e}pine, S., Hilton, E.~J., Mann, A.~W., et al.\ 2013, \aj, 145, 102 
%
\bibitem[Lissauer et al.(2014)]{liss14} Lissauer, J.~J., Marcy, G.~W., Bryson, S.~T., et al.\ 2014, \apj, 784, 44
%
\bibitem[Lissauer et al.(2011)]{liss11} Lissauer, J.~J., Ragozzine, D., Fabrycky, D.~C., et al.\ 2011, \apjs, 197, 8 
%
\bibitem[Mandel  \& Agol(2002)]{mandel} Mandel, K., \& Agol, E.\ 2002, \apjl, 580, L171
%
\bibitem[Mann et al.(2013)]{mann13} Mann, A.~W., Gaidos, E., \& Ansdell, M.\ 2013, \apj, 779, 188
%
\bibitem[Marcy et al.(2014)]{marcy13} Marcy, G.~W., Isaacson, H., Howard, A.~W., et al.\ 2014, \apjs, 210, 20
%
\bibitem[Muirhead et al.(2012)]{muir12} Muirhead, P.~S., Hamren, K., Schlawin, E., et al.\ 2012, \apjl, 750, L37
%
%
\bibitem[Petigura et al.(2013)]{petigura13} Petigura, E.~A., Howard, A.~W., \& Marcy, G.~W.\ 2013, Proceedings of the National Academy of Science, 110, 19273%
%
%
\bibitem[Pinsonneault et al.(2012)]{pins12} Pinsonneault, M.~H., An, D., Molenda-{\.Z}akowicz, J., et al.\ 2012, \apjs, 199, 30 
%
\bibitem[Press et al.(1986)]{numrec} Press, W.~H., Flannery, B.~P., \& Teukolsky, S.~A.\ 1986, Cambridge: University Press, 1986
%
\bibitem[Rowe et al.(2014)]{rowe14} Rowe, J.~F., Bryson, S.~T., Marcy, G.~W., et al.\ 2014, \apj, 784, 45 
%
\bibitem[Seager \& Mall{\'e}n-Ornelas(2003)]{seager03} Seager, S., \& Mall{\'e}n-Ornelas, G.\ 2003, \apj, 585, 1038
%
%
%
\bibitem[Silburt et al.(2015)]{silburt14} Silburt, A., Gaidos, 
E., \& Wu, Y.\ 2015, \apj, 799, 180 
{\bibitem[Still \& Barclay(2012)]{pyke} Still, M., \& Barclay, T.\ 2012, Astrophysics Source Code Library, 8004 }
%
%
{\bibitem[Torres et al.(2010)]{torres} Torres, G., Andersen, J., \& Gim{\'e}nez, A.\ 2010, \aapr, 18, 67}
%
\bibitem[VandenBerg et al.(2014a)]{vdbiso} VandenBerg, D.~A., Bergbusch, P.~A., \& Dowler, P.~D.\ 2014, Astrophysics Source Code Library, 4010
%
\bibitem[VandenBerg et al.(2014b)]{vdb} VandenBerg, D.~A., 
Bergbusch, P.~A., Ferguson, J.~W., \& Edvardsson, B.\ 2014, \apj, 794, 72 
%
%
\bibitem[Weiss 
\& Marcy(2014)]{weiss14} Weiss, L.~M., \& Marcy, G.~W.\ 2014, \apjl, 783, LL6 
%
%
\end{thebibliography}

\end{document}